\documentclass[superscriptaddress,showpacs,amsmath,amssymb,aps,prx,twocolumn,floatfix,nofootinbib]{revtex4-2}

\usepackage{graphicx}
\graphicspath{{figures/}}

\usepackage{dcolumn}
\usepackage{bm}
\usepackage{color}
\usepackage{amsmath}
\usepackage{amssymb}
\usepackage{bbm}
\usepackage{lineno}
\usepackage{hyperref}
\usepackage{hhline}
\usepackage{layouts}
\usepackage{color}
\hypersetup{
    colorlinks = true
}

\usepackage{natbib}
\usepackage{graphicx}
\usepackage{amssymb}
\usepackage{siunitx}

\usepackage[caption=false]{subfig}

\begin{document}

\title{Differentiable Simulation of a Liquid Argon Time Projection Chamber}

\newcommand{\SLAC}{SLAC National Accelerator Laboratory, Menlo Park, CA, 94025, USA}
\affiliation{\SLAC}
\newcommand{\APC}{Universit\'{e} Paris Cit\'{e}, CNRS, Astroparticule et Cosmologie, F-75013 Paris, France}
\affiliation{\APC}

\author{Sean~Gasiorowski} \email{sgaz@slac.stanford.edu}\affiliation{\SLAC}
\author{Yifan~Chen} \email{cyifan@slac.stanford.edu}\affiliation{\SLAC}
\author{Youssef~Nashed} \email{ynashed@slac.stanford.edu} \affiliation{\SLAC}
\author{Pierre~Granger} \affiliation{\APC}
\author{Camelia~Mironov} \affiliation{\APC}
\author{Daniel~Ratner} \affiliation{\SLAC}
\author{Kazuhiro~Terao} \affiliation{\SLAC}
\author{Ka~Vang~Tsang} \affiliation{\SLAC}

\newcommand{\comm}[1]{\textcolor{red}{#1}}
\newcommand{\corr}[1]{\textcolor{blue}{#1}}

\begin{abstract}
Liquid argon time projection chambers (LArTPCs) are widely used in particle detection for their tracking and calorimetric capabilities. 
The particle physics community actively builds and improves high-quality simulators for such detectors in order to develop physics analyses in a realistic setting. 
The fidelity of these simulators relative to real, measured data is limited by the modeling of the physical detectors used for data collection. This modeling can be improved by performing dedicated calibration measurements.
Conventional approaches calibrate individual detector parameters or processes one at a time. However, the impact of detector processes is entangled, making this a poor description of the underlying physics. We introduce a differentiable simulator that enables a gradient-based optimization, allowing for the first time a simultaneous calibration of all detector parameters. 
We describe the procedure of making a differentiable simulator, highlighting the challenges of retaining the physics quality of the standard, non-differentiable version while providing meaningful gradient information. We further discuss the advantages and drawbacks of using our differentiable simulator for calibration.
Finally, we provide a starting point for extensions to our approach, including applications of the differentiable simulator to physics analysis pipelines. 
\end{abstract}

\keywords{differentiable simulator}

\maketitle
\tableofcontents

\section{Introduction}
High quality simulations of physics detectors are a fundamental piece of infrastructure across a diverse set of scientific disciplines, with uses ranging anywhere from physical inference to experimental design. 
In high energy particle physics measurements, such as those targeted by DUNE~\cite{DUNE:2020ypp}, detector simulation is particularly crucial to analysis and event reconstruction.
However, detector simulations often have discrepancies relative to the real experiments, which can lead to systematic errors and biased results, reducing the sensitivity and accuracy of physics measurements. Dedicated calibration procedures are therefore required to minimize the differences between these two regimes.

In particle physics, a conventional approach for identifying sources of data-simulation difference is to isolate different detector modeling processes using selected control samples~\cite{MicroBooNE:2019efx, DUNE:2020cqd, MicroBooNE:2021icu, Box, Birks, MicroBooNE:2019koz, MicroBooNE:2020kca}. These control samples are auxiliary datasets designed to amplify the impact of particular detector effects. Calibration using these control samples may proceed by either directly altering the simulation of the amplified underlying physics process or by performing an \emph{ad hoc} correction on observables without direct modification of the simulation. 
However, different detector processes can affect the measured data in similar ways, meaning that an isolated approach may not capture the interplay among entangled detector processes, making it difficult to fully address sources of data-simulation difference. The \emph{ad hoc} approach~\cite{MicroBooNE:2019efx}, by not including explicit correction of the physics models, may fail to capture upstream effects. 

A simultaneous correction of all detector processes and physics models would be an ideal new paradigm for detector calibration. However, this has not been previously achieved due to limitations of the existing software tools and calibration frameworks in addressing the potentially high dimensionality of the associated parameter space.

Gradient-based optimization provides a pathway towards improving calibration. Optimizers based on gradients are the cornerstone of modern machine learning, used in the training of many of the most common machine learning methods, notably neural networks~\cite{DeepLearning}. Gradient-based methods are powerful and efficient in high dimensional optimization, supporting a simultaneous fitting of arbitrarily many parameters. Equipping existing detector simulations with gradient information enables the use of gradient-based optimization for calibration.
Further, as gradients are used to fit a set of physical parameters within the same detector simulation used for physics analysis, application of this calibration is trivial -- the fitted parameters are simply used within the same simulation code. Therefore, calibration results can be easily tracked and consistently applied. By adjusting physical model parameters, this calibration also provides a characterization of the detector for direct experimental feedback. 

Gradient-based optimization requires efficient calculation of the necessary gradients. A variety of software packages, such as PyTorch~\cite{PyTorch}, TensorFlow~\cite{Tensorflow}, and JAX~\cite{JAX}, are capable of performing this calculation. These tools are commonly used to support the training of neural networks, in which gradients of a loss function with respect to neural network parameters are used for a minimization. The calculation of these gradients is done using a set of techniques called automatic differentiation~\cite{AutoDiff}. In automatic differentiation, computer code is decomposed into a set of fundamental operations with known derivatives. Derivatives of the full program may then be calculated using these fundamental operations and the chain rule. As automatic differentiation is implemented in these machine learning software packages, any code written using these frameworks can use automatic differentiation for calculating gradients, including physics simulation. This type of approach is broadly called differentiable programming.

Methods other than gradient-based optimization using automatic differentiation exist for solving the problem of multi-dimensional parameter fitting. Approximate gradient methods~\cite{ApproximateGradient}, e.g. based on finite difference approximations to the gradient, do not require rewriting code into an automatic differentiation framework. However, such methods become computationally intractable in terms of number of function evaluations as the number of dimensions increases, and they rely on numerical choices such as step size between finite difference function evaluations, which may introduce inaccuracies.

Evolutionary or genetic algorithms~\cite{GeneticAlgorithms} are popular for black-box optimization, and require no rewriting of the simulator or any assumptions about differentiability. However such models rely on explicit exploration of the search space, which becomes exponentially large as the number of dimensions increases, leading to slow or poor convergence. Bayesian optimization~\cite{BayesianOpt} describes another class of black box algorithms, but requires additional assumptions, e.g. the choice of a prior and a surrogate model. The fit quality and computational complexity of this surrogate (often a Gaussian process) also often scales poorly due to the exponentially large search space as the number of fitting dimensions increases.

Gradient-based optimization using automatic differentiation requires adapting simulation code into an automatic differentiation framework. It further assumes differentiability and may become trapped in local optima in the optimization landscape. However, unlike approximate gradient methods, automatic differentiation is computationally efficient in high dimensions, with a cost incurred only by an increased number of stored intermediate results. The gradients computed by automatic differentiation are exact, avoiding numerical choices such as finite difference step size. Gradient-based optimization is able to achieve convergence to optima in high dimensional spaces; it is used for training even the largest neural networks, which can have billions of parameters, an infeasible task for genetic algorithms or Bayesian optimization. Furthermore, gradient-based optimization aids in direct interpretation of results, as gradients are directly tied to simulation output, and iterations are sequential steps in optimization parameters. 

We would like to demonstrate the utility of gradient-based optimization for a detector calibration task. Liquid argon time projection chambers (LArTPCs) are an excellent candidate for this demonstration. LArTPCs are used in a variety of modern particle physics experiments. They have several appealing experimental qualities~\cite{Rubbia:1977zz}, and are capable of measuring both particle trajectory and energy, known as tracking and calorimetry respectively. DUNE~\cite{DUNE:2020ypp} is a developing neutrino experiment which relies heavily on LArTPCs. There is significant community and governmental investment in DUNE, meaning that there is much interest and effort behind high quality LArTPC simulation targeting neutrino physics, as well as significant potential impact in improving the corresponding calibration pipeline. These factors, as well as the conceptual simplicity of the detectors, make LArTPCs a prime target for developing a gradient-based calibration.

We therefore present a differentiable simulator for a DUNE LArTPC prototype. Starting from a snapshot of the simulator presented in Ref.~\cite{larnd-sim}, we have made a variety of modifications to allow for the calculation of gradients via automatic differentiation. These gradients explicitly describe the dependence of simulator predictions on simulation parameters, and provide a natural, efficient, and automatic path towards detector calibration.

The physical models used in simulation may be approximate descriptions of detector processes, and there may be additional components not captured by these models. This means that after a calibration of parameters of these models, there may still be discrepancies between simulation and data. The differentiable simulator specifically targets the improvement of parameter-based calibration. Extensions to account for missing or incomplete simulation components are left for future work.

In this paper we describe the steps taken to go from a conventional to a differentiable LArTPC simulation, demonstrate the capabilities of the simulator for the calibration of simulation parameters, and analyze the computational bottlenecks and pitfalls to be addressed for broader adoption of this tool.


\section{Simulator implementation}
\label{sec:sim-description}

A LArTPC is a detector with an applied electric field across a volume of liquid argon. 
Deposition of energy can cause ionization in the liquid argon, and the ionization electrons drift under the influence of the electric field to be measured as current in an electronics readout. 
Our simulator uses a detector configuration corresponding to a module design for a DUNE LArTPC prototype. The rectangular detector module has a size of $\SI{60}{\cm}\times\SI{60}{\cm}\times\SI{120}{\cm}$, and contains two back-to-back LArTPCs. The electric field in each LArTPC is generated by an applied voltage difference between each anode and a cathode plane. The cathode plane is shared and located between the two LArTPCs. The two anode planes are parallel to the cathode plane on respective edges of the detector volume.  Both anode planes consist of a pixelated charge readout called LArPix~\cite{Dwyer:2018phu}. We define $x$ and $y$ as the horizontal and vertical axes along the cathode plane, with $z$ corresponding to the drift axis. Resolution of the LArPix readout in each dimension is determined by the pixel pitch and time sampling frequency. The LArPix pixel pitch is \SI{4.4}{\mm}. The spatial resolution in the drift direction is below \SI{2}{\mm}.

The simulator developed for the DUNE LArTPC prototype is called \texttt{larnd-sim}~\cite{larnd-sim}. Energy depositions are represented as line segments (``particle segments'') defined by 3D start and end positions, $(x, y, z)$, and an associated energy, $E$.
The simulator models the production of ionization electrons due to these energy depositions, as well as the drifting of electrons to the pixel readout at each anode plane.
A current in the pixel readout can be induced by charges approaching or being directly collected on a given pixel.
Each pixel is an independent readout channel with its own setup for trigger threshold and gain.
Currents in the pixel readout are digitized to analog-to-digital converter (ADC) counts before being read out. The location of induced current in the pixel plane provides an $x,y$ measurement. Timing information provides information on $z$. Magnitude of the induced current (ADC counts) gives information on deposited energy.

\begin{figure}
  \centering
  \includegraphics[width=\linewidth]{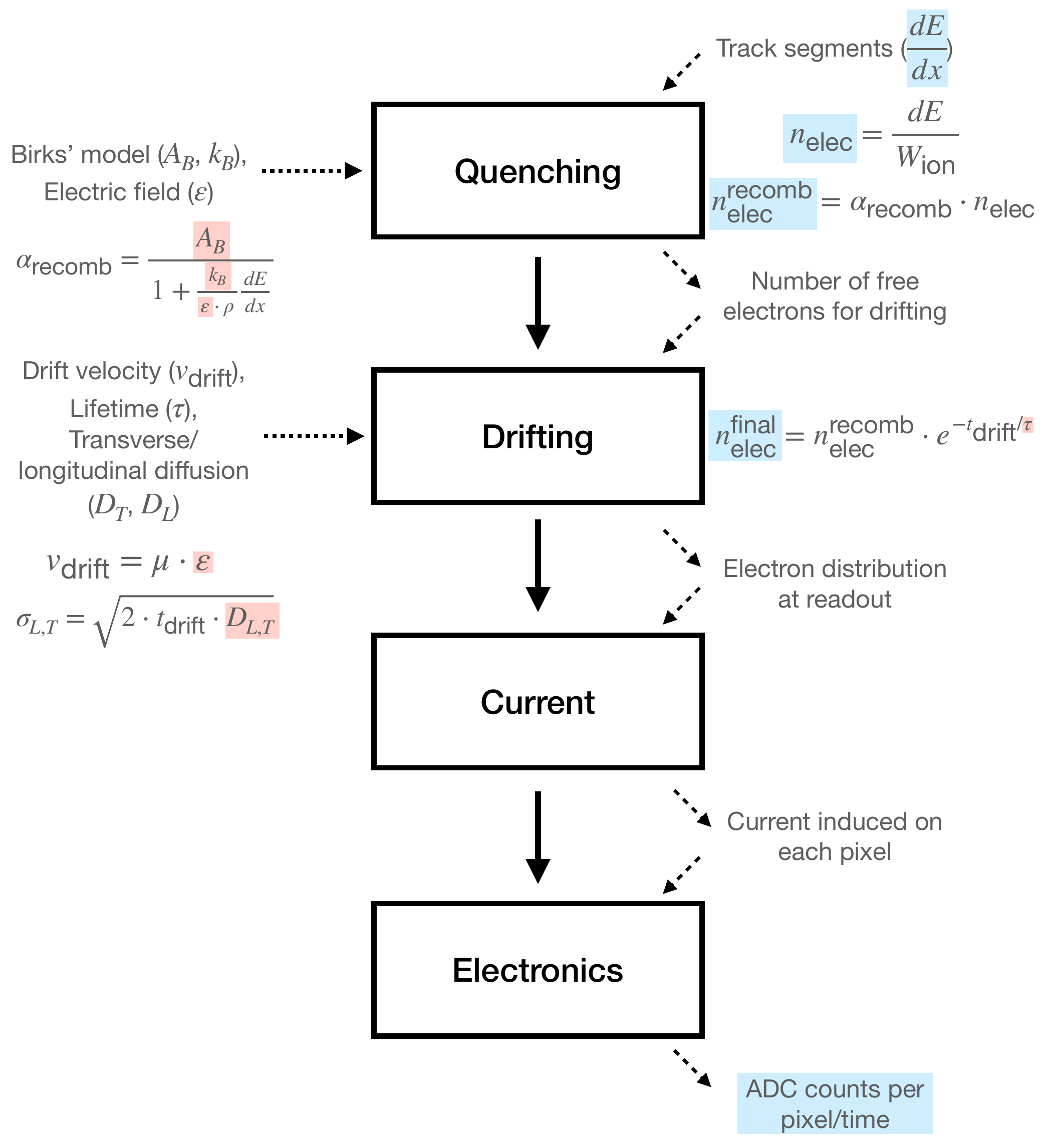}
  \caption{
   Flow diagram of the simulator, highlighting inputs and outputs of each stage (blue) as well as commonly calibrated model parameters (red).
   }
  \label{fig:simulation-flow}
\end{figure}

We describe the details of the simulator in the following stages corresponding to sequential physics processes: 1) Charge quenching; 2) Electron drifting; 3) Current accumulation; 4) Electronics simulation.
We highlight several of the physics models used in the simulator. A summary of information about the simulator is presented in Fig.~\ref{fig:simulation-flow}.

\textbf{Charge quenching.}
In this stage, the simulator determines the number of ionization electrons produced given the energy deposition per particle segment, $dE$. This number is given by
\begin{equation}
n_{\text{elec}} = \frac{dE}{W_{\text{ion}}},
\end{equation}
where $W_{ion} = \SI{23.6}{\eV}$ is the work function, i.e., the average energy required to produce an ionized electron in liquid argon~\cite{Ionization-Energy, Models}.

The ionized electrons may subsequently recombine with nearby argon ions.
The electron survival rate, describing the fraction of electrons which do not recombine, depends on the applied electric field and the local charge density.
There are two commonly used recombination models in modern LArTPCs: the Birks model~\cite{Birks} and the modified box model~\cite{Box}, which yield similar results for our uses.
For this work, we primarily use the Birks model to describe the process of electron recombination, and the electron survival rate takes the form
\begin{equation}
\alpha_{\text{recomb}} = \frac{A_{B}}{1+\frac{k_{B}}{\mathcal{E}\cdot\rho}\frac{dE}{dx}},
\label{eq:Birks}
\end{equation}
where $\mathcal{E}$ is the applied electric field, $\frac{dE}{dx}$ is the energy deposition per unit length, which gives the local charge density, and $\rho$ is the liquid argon density.
$A_{B}$ and $k_{B}$ are parameters of the Birks model and are typically fit in conventional calibration procedures.

After recombination, the number of electrons is
\begin{equation}
n_{\text{elec}}^{\text{recomb}} = \alpha_{\text{recomb}} \cdot \frac{dE}{W_{\text{ion}}}.
\end{equation}

\textbf{Electron drifting.}
Energy depositions in the simulator are represented using 3D spatial coordinates, $(x,y,z)$. For simulation of the readout, the position, $z$, along the drift axis must be translated into a \emph{drift time}, $t_{\text{drift}}$.
The electric field in the simulator is considered to be uniform and perpendicular to the readout planes. Therefore, the relationship between drift time and distance from a given anode along the drift axis is given by $\Delta z = v_{\text{drift}} \cdot t_{\text{drift}}$.
In the simulation, the \emph{drift velocity}, $v_{\text{drift}}$, is directly associated to the electric field, $\mathcal{E}$, and is given by 
\begin{equation}
v_{\text{drift}} = \mu(\mathcal{E}, T)\cdot \mathcal{E}.
\end{equation}
The electron mobility, $\mu(\mathcal{E}, T)$, is a function of electric field, $\mathcal{E}$, and liquid argon temperature, $T$, and is an empirical model derived from previous measurements~\cite{vdrift-param}.
The simulated $t_{\text{drift}}$ is then calculated as the drift distance divided by $v_{\text{drift}}$.

On their path to the readout, the ionized electrons that have survived recombination can be further consumed by electronegative impurities, such as oxygen and water~\cite{Lifetime}.
Assuming the impurities are distributed uniformly in the liquid argon, the survival rate depends on $t_{\text{drift}}$ and the impurity type and level, which can be summarized by the electron lifetime $\tau$, the mean drift time of an electron before capture by an electronegative impurity.
The number of electrons that arrive to the readout after both recombination and drifting is therefore modeled by
\begin{equation}
n_{\text{elec}}^{\text{final}} = n_{\text{elec}}^{\text{recomb}} \cdot e^{-t_{\text{drift}}/\tau}.
\end{equation}

During the drifting stage, we also calculate charge diffusion, describing the spread of the ionization electrons in 3D space, which depends on $t_{\text{drift}}$.
The corresponding longitudinal and transverse diffusion lengths, $\sigma_{L}$ and $\sigma_{T}$, determine the scale of the charge spread along the drift direction ($z$) and in the pixel plane ($x,y$) respectively. The resulting distribution is an important input for the simulation of the readout.

Longitudinal and transverse diffusion lengths are given by
\begin{equation}
\sigma_{L, T} = \sqrt{2\cdot t_{\text{drift}} \cdot D_{L, T}},
\end{equation}
where $D_{L}$ and $D_{T}$ are longitudinal and transverse diffusion coefficients in liquid argon. These coefficients are usually extracted via separate, targeted measurements.

\textbf{Current accumulation.}
In this stage, we model the distribution of charge on the pixel readout for each given particle segment and use that distribution to calculate the generated current on the appropriate pixels. The total charge for a given segment is determined by the number of surviving ionization electrons. This is assumed to be spread along the length of the segment, and the impact of diffusion is modeled via a 3D Gaussian with widths given by the diffusion lengths $\sigma_{L, T}$ for longitudinal ($z$) and transverse ($x$ and $y$) components relative to the drift direction.
The charge distribution is convolved with a functional current model, which depends on the electron drift time and the pixel position.
Integration is done by summing the contributions from points which are sampled in a cube surrounding each particle segment.
The current integration is one of the most computationally intensive steps of the simulation.
See Sec.~\ref{subsec:computation} for more details.

\textbf{Electronics simulation.}
The final stage of the simulation models the charge readout electronics. It incorporates the calculation of signals on relevant pixels, including the contribution from different particle segments, as well as triggering and digitization of current into ADC counts.
The LArPix readout is nominally configured so that each pixel is a separate readout channel. These readout channels have a sampling rate of \SI{10}{\mega\Hz}.
Pixels in the readout are ``self-triggering'', meaning that they continuously take data. If the integrated current passes a given trigger threshold, the corresponding readout channels will integrate the current over a window of \SI{1.8}{\micro\second} and record the result.
The time when the signal first rises above the threshold is registered as the time of the corresponding pixel ``hit''. Readout channels are reset after this signal integration.

Stochasticity in \texttt{larnd-sim} is introduced to model noise in the LArPix electronics system. The simulator includes two types of noise: (1) a baseline uncorrelated noise, e.g. from thermal fluctuation, and (2) noise that occurs during the readout channel reset. Both sources of noise are additive with respect to the pixel current integration and are drawn from Gaussian distributions, each with mean at $0$ and standard deviation set according to LArPix measurements.
In this work, results are presented with both noise models turned off. See Sec.~\ref{sec:param-fitting} for a more detailed discussion of this choice.

\subsection{Software infrastructure overview}
\label{subsec:software_overview}
The primary implementation effort for this work has been focused on translating a snapshot of \texttt{larnd-sim}, based on CUDA kernels written using Numba~\cite{Numba}, into an automatic differentiation framework~\cite{AutoDiff}. The framework chosen for this rewriting is EagerPy~\cite{EagerPy}, designed to be agnostic to a user's choice of automatic differentiation backend (e.g. JAX, TensorFlow, or PyTorch), allowing flexibility in the particular choices of users building applications on top of the differentiable simulation. Usage of EagerPy code does require a choice of framework, and we use PyTorch for the results presented here.

Many of Python's automatic differentiation frameworks are designed efficiently around matrix and tensor operations, with near trivial GPU acceleration of such operations. Explicit CUDA kernels, on the other hand, are phrased in terms of individual operations on given GPU threads. In order to take advantage of the efficiency of these matrix and tensor operations, we have therefore \emph{vectorized} the \texttt{larnd-sim} snapshot where possible. 
Writing the simulation in such a way allows for GPU acceleration by merely specifying the device of the simulation inputs, and takes advantage of the optimized tensor operations of these automatic differentiation tools. However, this comes at the cost of higher memory usage due to large tensor sizes.

\subsection{Differentiable relaxations considered}
\label{subsec:relaxations}
Developing a differentiable LArTPC simulator requires the assumption of differentiability of all operations that depend on the variables of interest. As gradients are calculated using the chain rule, this sequence of operations runs from the introduction of the relevant variables up until the simulator output, but might not include all operations within the simulator. Even if all operations in the sequence are differentiable, cases arise where gradients either vanish or are infinite, meaning that such gradients exist but are not useful for applications such as gradient-based fitting. One way of avoiding such issues is to introduce a set of \emph{differentiable relaxations}: continuous, often smooth approximations of non-differentiable or poorly conditioned operations. Two scenarios of particular relevance for this work which require such relaxations are (1) discrete integer operations and (2) hard masking operations. 

\begin{enumerate}
\item In our simulator, several quantities are discrete (e.g. number of electrons, ADC counts). In practice, such quantities result from floating point operations which are discretized, e.g. via truncation. This poses an issue for the gradient calculation in two ways. Recall that a derivative is formulated as 
\begin{equation}
f'(x) = \lim_{h\rightarrow 0} \frac{f(x+h) - f(x)}{h}.
\end{equation}
In the case of truncation, there is a range of input $x$ values for which the truncated output produces the same result (e.g., for truncation to integer values, $\text{trunc}(x)$ = 0 for all $x\in [0, 1)$). This means that the gradient is 0. Additionally, the transition between discrete output values is of a step function nature, with an undefined (infinite) derivative at the transition point (e.g. $\text{trunc}(1) = 1$). All resulting gradients are therefore 0 or infinite, leading to poor performance of a gradient-based optimization.

To remedy this, we apply a differentiable relaxation, in this case by removing the truncation, allowing for a continuous variation of these nominally discrete values. The continuous variation allows for nonzero gradients, providing useful gradient information for downstream use. This relaxation is only necessary during any gradient-based optimization, and truncation may be applied after the optimization is complete to restore the physical meaning of the discrete quantities.

\item Present throughout the simulator as well are a variety of masking operations based on cuts, e.g. for variable $x$, requiring $x>a$ for some fixed $a\in\mathbb{R}$. If the quantity $\frac{df(x)}{dx}$ is desired, this poses the same issue as discussed above, again due to the step function nature of such a cut. Depending on the desired application, derivatives through all such masks may not be required. However, one notable example for our particular case is a truncated exponential, defined as
\begin{equation}
f(x) =
    \begin{cases}
        e^{-x} & x > 0\\
        0 & x \leq 0
    \end{cases}
\end{equation}
which appears as part of the current model. To aid in the gradient computation, the hard mask of $x > 0$ is replaced by a sigmoid function, defined as
\begin{equation}
    \text{Sigmoid(x, s)} = \frac{1}{1+e^{-k\cdot x}}.
\end{equation}

\begin{figure}
  \centering
  \includegraphics[width=0.8\linewidth]{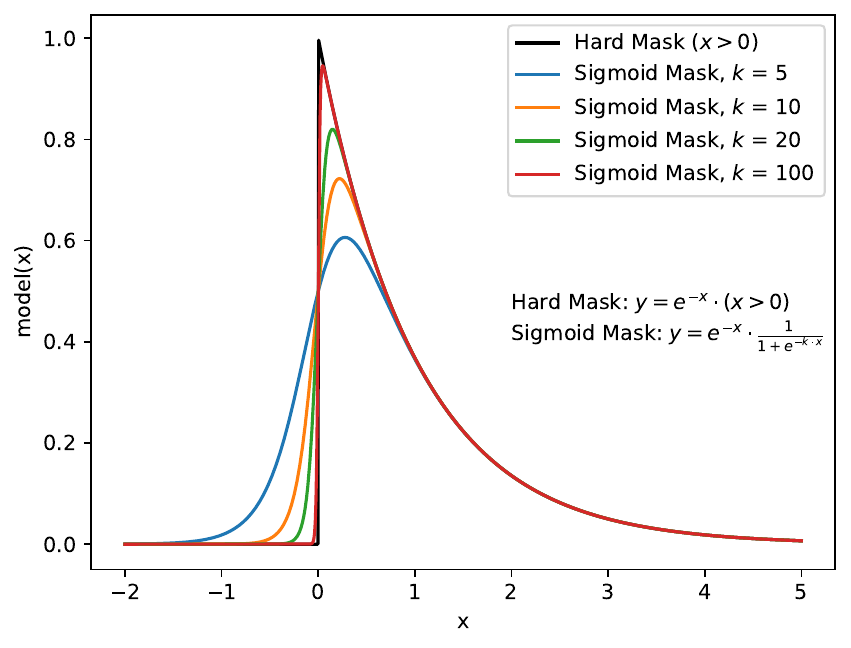}
  \caption{Sigmoid masking of the truncated exponential function for various values of rate, $k$. 
}
  \label{fig:sigmoid-mask}
\end{figure}

This function smoothly transitions from $0$ to $1$ as $x$ moves from $-\infty$ to $+\infty$, with the rate of transition controlled by parameter $k$, approaching a step function as $k\rightarrow \infty$. A demonstration of how the approximate truncated exponential model changes with $k$ is shown in Fig.~\ref{fig:sigmoid-mask}. Higher values of $k$ correspond to a closer match with the unsmoothed truncated exponential, and $k$ may be tuned to trade-off between accuracy and gradient performance. A value of $k=100$ is used in the results presented here. In the latest version of \texttt{larnd-sim}, the current model has been replaced by a lookup table, which itself poses some challenges for differentiability. However, such a structure may be incorporated into a differentiable framework via a neural network parametrization, as in Ref.~\cite{SIREN}.
\end{enumerate}

The pixel readout, represented by coordinates $x$ and $y$, is itself a discrete system. Measurement of timing ($t$) is also done as a discrete sampling at fixed intervals. This discreteness is challenging in our simulator, as $x$, $y$, and $t$ are represented via matrix and tensor indices, rather than standalone quantities. Derivatives with respect to $x$, $y$, and $t$ reflect changes in simulated quantities across pixels and times. For the application considered here, these changes do not have a large impact, and the corresponding derivatives can be safely ignored. 
For applications beyond this paper, we have implemented a differentiable relaxation of an important use of timing information in the trigger logic, which we describe below. Differentiable relaxations of pixel systems may also be developed, which can be considered for future work.

\begin{figure}
  \centering
  \includegraphics[width=0.8\linewidth]{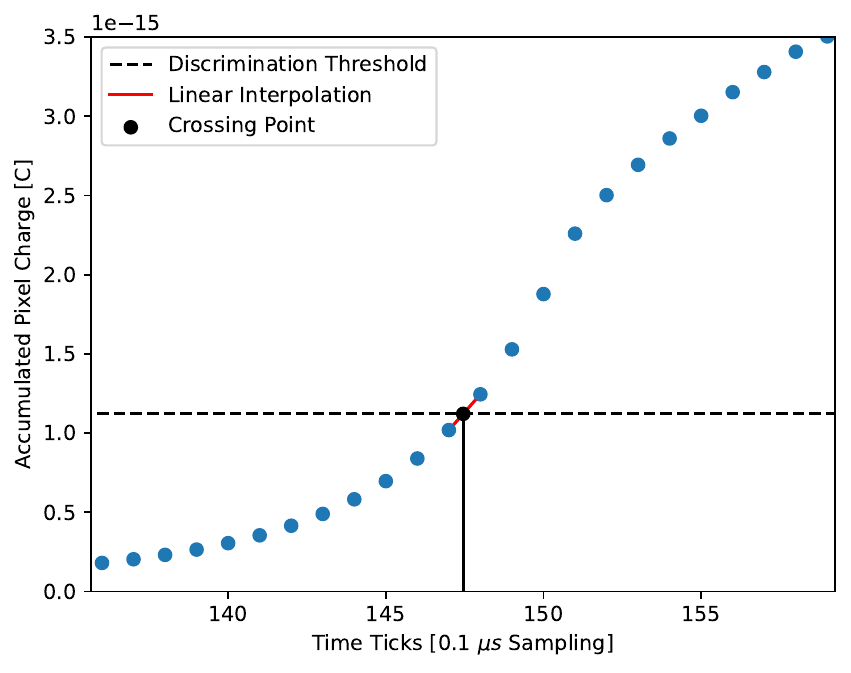}
  \caption{Illustration of the interpolated trigger timing index procedure for a single pixel.
  }
  \label{fig:trigger-demo}
\end{figure}

For LArPix triggering, given some cumulative charge as a function of discrete time sampling indices, we need to determine the time at which the cumulative charge crosses a particular value threshold. As this time is stored via an integer index, we have both the same discrete problem as above, and also a framework problem, as indices must be integers, which breaks the gradient flow. To remedy this, we first recognize that the gradient only depends on the local neighborhood of the threshold intersection. We may therefore (1) find the two discrete points surrounding the intersection, (2) linearly interpolate to find a continuous time of intersection, and then (3) store this continuous value as the relevant timing information. Fig.~\ref{fig:trigger-demo} illustrates this procedure on a single pixel.

The simulated output using all of these relaxations was compared with a reference simulation from our snapshot of \texttt{larnd-sim}. Both simulators produced very similar results, with an average deviation of 0.04 ADC counts per activated pixel, which is two orders of magnitude below the typical noise level for the input dataset defined in Sec.~\ref{sec:data}.

\subsection{Computational performance}
\label{subsec:computation}
To support the use of our differentiable simulator by analyzers, we provide a detailed assessment of its computational demands, both in isolation and relative to our non-differentiable snapshot of \texttt{larnd-sim}. We assess performance of our differentiable simulator by analyzing two main metrics: computation time and GPU memory usage. All analysis is done using single NVIDIA Tesla A100 GPUs on the Scientific Data Facility at SLAC. 

When using our differentiable simulator for calibration, there is a cost incurred by both the forward simulation and the gradient computation, which is done via reverse mode automatic differentiation~\cite{AutoDiff}, also known as \emph{backpropagation}. We present metrics relevant to both the forward simulation in isolation and the forward simulation together with backpropagation. Comparison to \texttt{larnd-sim} is done with the forward simulation only.

\subsubsection{Computation time}
\begin{figure}
    \centering
    \includegraphics[width=\columnwidth]{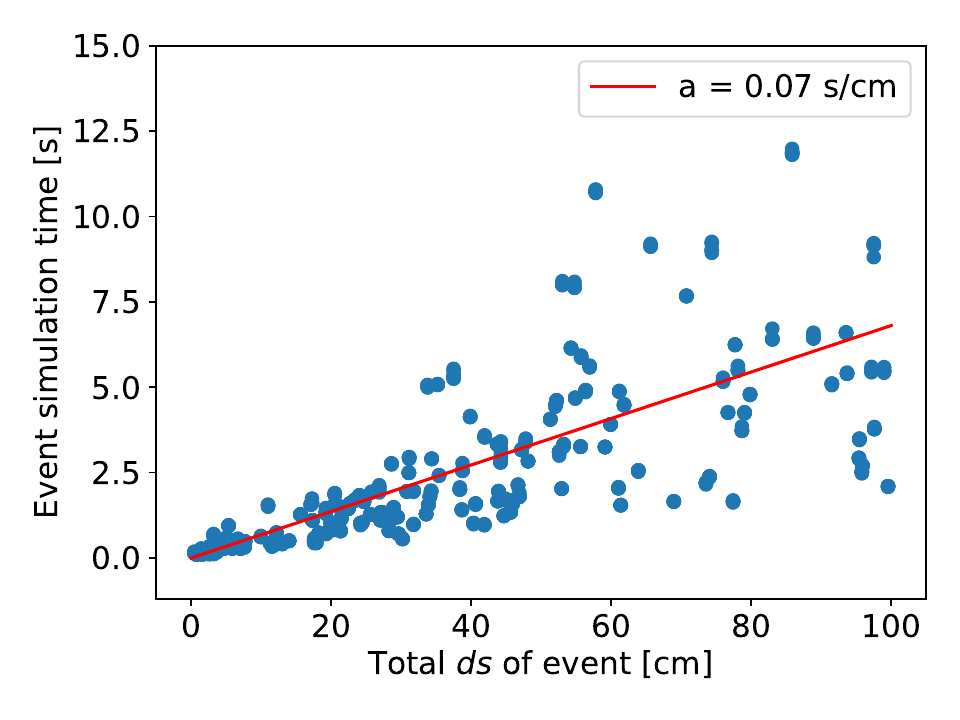}
    \caption{
    Simulation time of events as a function of $ds$, the total segment length.
    The red line shows a linear fit to the data.}
    \label{fig:time_vs_dx}
\end{figure}

We discuss the computation time of the differentiable simulator in terms of simulation time, which is the cost of the forward simulation in isolation, and fitting duration, which includes both the forward simulation and backpropagation. 
Simulation time changes between events, depending on a variety of interdependent variables such as number of particle segments in the event, the length of each of those segments, and the number of pixels with current induced. We found that this information may be summarized by the total segment length ($ds$), defined as the sum of the lengths of a collection of particle segments (e.g. an event or a batch, see Sec.~\ref{subsec:procedure}).
This is demonstrated by the clear correlation between $ds$ for each event and simulation time shown in Fig.~\ref{fig:time_vs_dx}. The average duration per event $ds$ is \SI{0.07}{\second\per\centi\meter}. 

\begin{figure}
    \centering
    \includegraphics[width=\columnwidth]{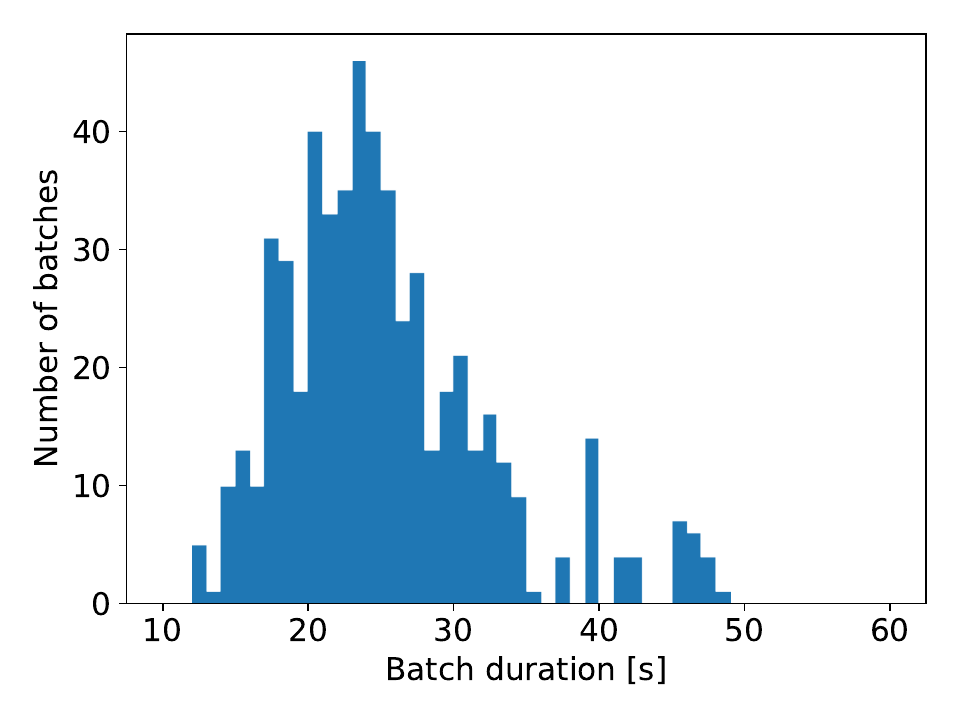}
    \caption{Distribution of the processing time (simulation and backpropagation) for batches of $ds= \SI{100}{\cm}$.
    }
    \label{fig:batch_duration}
\end{figure}

The correlation between $ds$ and simulation time motivates the use of this variable for batching the events during a gradient-based fit (Sec.~\ref{sec:param-fitting}), as such a batching balances the computation time required for fitting each batch, which is useful for both large scale computation and in setting the stage for future parallelization (e.g. with multiple GPUs). The distribution of the batch fitting durations is presented in Fig.~\ref{fig:batch_duration} for a batch size of $ds = \SI{100}{\centi\meter}$. The resulting distribution is grouped around \SI{25}{s}. This duration includes both the forward simulation and the gradient computation.

We have also tested the impact of different batch sizes, determined by $ds$ values, on the \emph{total} computation time of fitting a given dataset with a set number of epochs.
While the computation time per fitting iteration is generally correlated to the batch size, the batch size has minimal impact on the computation time of the overall fit.

Because the gradient computation involves the storage of a large number of intermediate results, memory usage quickly becomes infeasible for desirable batch sizes (such as for $ds=\SI{100}{\cm}$) on single A100 GPUs. Therefore, we take advantage of PyTorch \emph{checkpointing}, which discards these intermediate results and recomputes them during gradient accumulation. This reduces memory usage at the cost of increased computation time. The impact of checkpointing is included in Fig.~\ref{fig:batch_duration}.

\begin{figure}
    \centering
    \includegraphics[width=0.95\columnwidth]{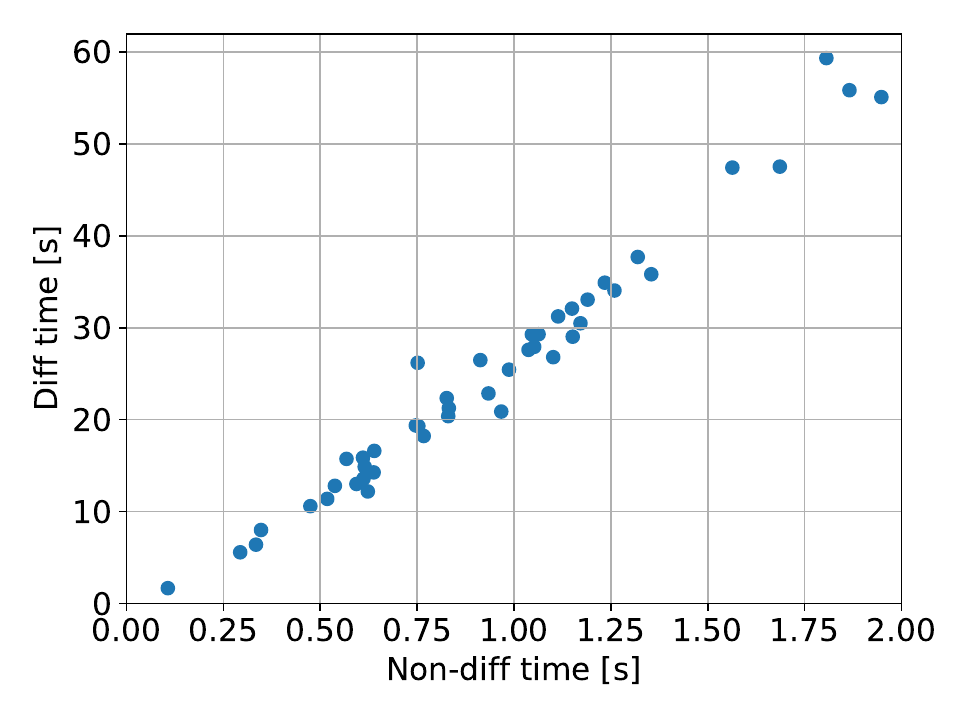}
    \caption{A comparison of the simulation time per event between the original \texttt{larnd-sim} and the differentiable version.
    }
    \label{fig:diff_time_compare}
\end{figure}

To assess the performance difference between \texttt{larnd-sim} and our differentiable rewrite, we process an identical sample of events through both our snapshot of \texttt{larnd-sim} and our differentiable simulator and compare the simulation time event-by-event.
Fig.~\ref{fig:diff_time_compare} shows that the simulation time of the differentiable simulator is roughly 25 times longer than the non-differentiable snapshot.
Importantly, this comparison is done without the gradient computation for the differentiable simulator. This is a significant performance gap. However, there are a few concrete areas for further improvement.

The first area of improvement is related to the vectorization mentioned in Sec.~\ref{subsec:software_overview}. This vectorization is necessary for the use of automatic differentiation frameworks such as PyTorch. However, it tends to increase the total number of operations, as all operations are framed as large dense matrices instead of dedicated, element-wise kernels.
LArTPC simulation is often a sparse problem, as typically readout signals per event only take place in a small fraction of the detector volume. Therefore, the increase in operations is expected to add significantly to the additional computational overhead. Alternative automatic differentiation tools, such as Enzyme~\cite{Enzyme} may be able to avoid this vectorized rewrite, and can be considered for future computational development.

Furthermore, \texttt{larnd-sim} is just-in-time (JIT) compiled~\cite{Numba}. This compilation avoids the overhead of interpreted code, and is expected to significantly decrease computation time. Although PyTorch does have JIT compilation tools, introduction of JIT compilation would require significant additional study, and therefore was not done for this first demonstration.

Because of the deep integration of JIT compilation with Numba's CUDA library, separation of the effects of vectorization and JIT compilation is not achievable without a significant alteration of either the original \texttt{larnd-sim} snapshot or our differentiable version. 
Therefore, the performance gap between the two versions includes the combined effects of both sparse computation and JIT compilation.

\subsubsection{Memory usage}
In nearly every large scale computation, there is a trade-off between memory usage and computation time. Understanding memory usage in our application and being able to estimate it in advance allows us to do specific optimizations aimed at efficiently using allotted memory resources, thereby reducing computation time.

Peak memory usage in our simulator occurs during the computation of the signal on each pixel (the \textbf{Current accumulation} stage of the simulation), which is also the most computationally intensive step of our snapshot of \texttt{larnd-sim}~\cite{larnd-sim}. In our case, the high memory usage is due to the construction of a very large, dense matrix. Given a particle segment and a pixel, the signal has to be computed by sampling in four dimensions: parallel to the readout planes, along the $x$ (1) and $y$ (2) axes; across time of arrival at the anode (3); across time of evaluation of the current (4). We denote the number of samplings in each of these four dimensions as $N_x$, $N_y$, $N_{T_0}$, and $N_{T_f}$ respectively. 
The spacing between sampled points in $x$ and $y$ is identical for all pixels, with $N_x = N_y = 30$ points per pixel. 
The time samplings, however, are data dependent, and the following formulas give upper bounds on the required tensor dimensions:

\begin{equation}
    N_{T_0} \leq \underset{\forall \text{seg}}{\max}{\left|\max{(d, 5\sqrt{2}\sigma_T)}\sqrt{1+\cot^2{\phi}} + 4\sigma_L\right|}\times\frac{4}{T_\text{sampling}}
\label{eq:NT0}
\end{equation}

and

\begin{equation}
    N_{T_f} \leq \frac{\underset{\forall \text{seg}}{\max}\left|\Delta_z\right| + 0.5 + 2T_\text{padding}\times v_\text{drift}}{T_\text{sampling}\times v_\text{drift}} + 1,
\end{equation}
with $d$ as the diagonal length of the pixel pad, $\sigma_T$ as the transverse diffusion distance, $\phi$ as the angle of the particle segment relative to the drift direction, $\sigma_L$ as the longitudinal diffusion distance, $T_\text{sampling}$ as the LArPix sampling rate, $\Delta_z$ as the $z$ range of the segment, and $T_\text{padding}$ as the number of recorded samples before (after) the start (end) of signal on the pad.

Eq.~\ref{eq:NT0} shows that $N_{T_0}$ depends on angle $\phi$.
When $\phi$ approaches 0, $\cot\phi$ goes to infinity. This means that $N_{T_0}$, and therefore the memory usage, becomes unbounded for particle segments parallel to the drift direction. A simple angular cut can limit the memory usage, and is applied in the calibration data set (see Sec.~\ref{sec:data}).

\begin{figure}
    \centering
    \includegraphics[width=0.95\columnwidth]{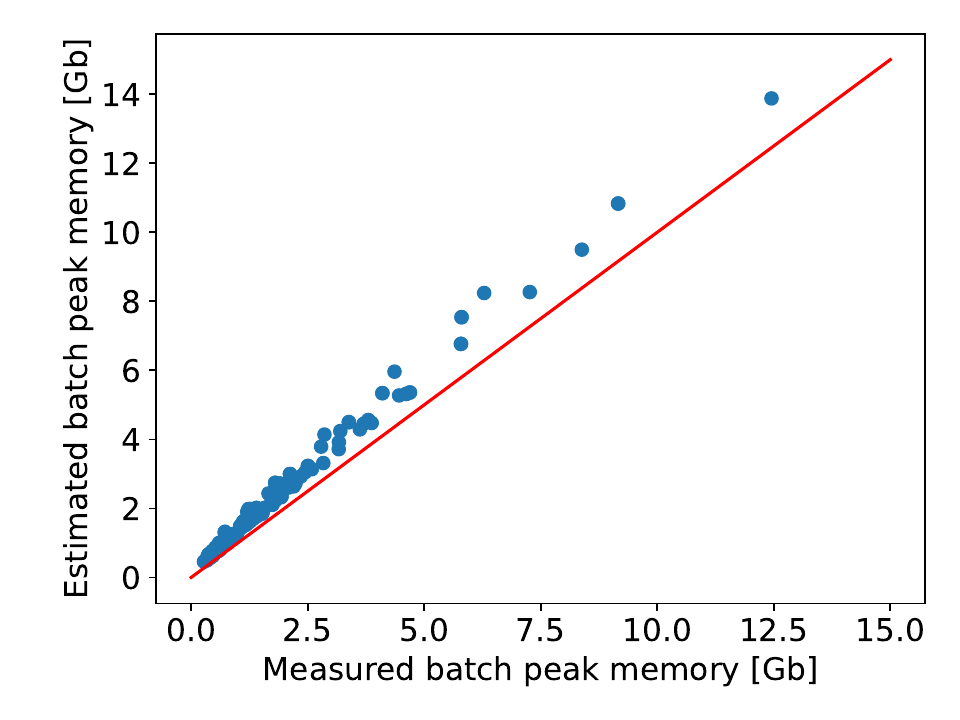}
    \caption{The estimated batch peak memory use in comparison with the measured batch peak memory. The red line shows the equality relation. Points being above it mean that the measured memory is consistently smaller than the estimated one.}
    \label{fig:mem_upper_bound}
\end{figure}

The size of the tensor in this peak memory estimation is given by $\mathcal{M} = N_\text{segments} \times N_\text{pixels} \times N_{T_f} \times N_{T_0} \times N_x \times N_y$. Two copies of this tensor are used in the most memory intensive computation.
The estimated peak memory is therefore 2 times the tensor size multiplied by the memory used per tensor element (32 bits for floating point precision).
Fig.~\ref{fig:mem_upper_bound} shows that the estimated peak memory per batch is very close to the measured value and that all measured peak memory values fall below the estimated upper bound.

Although the current calculation is vectorized, this peak memory stage of the simulation is converted into a loop over particle segments and pixel ``chunks'', which are batched subsets of the full pixel set. The pixel chunks are sequentially passed through the memory intensive computation. This chunking operation directly trades-off between memory usage and computation time. A larger chunk size leads to higher memory usage, but requires fewer loop iterations, and thus takes less time.

\begin{figure}
    \centering
    \includegraphics[width=0.95\columnwidth]{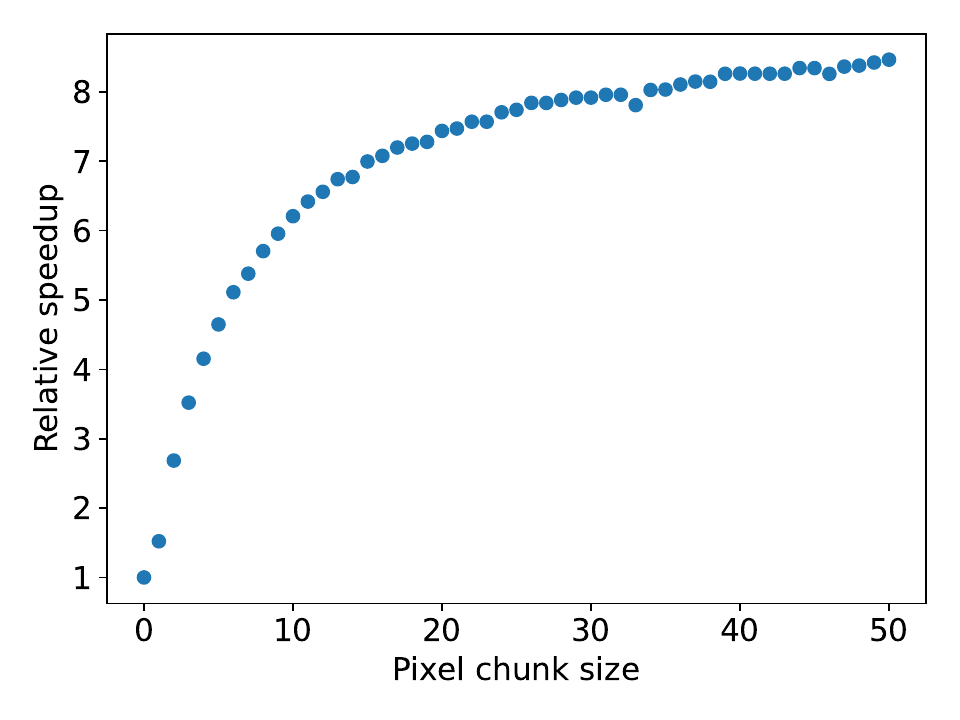}
    \caption{The relative speed-up factor as a function of the pixel chunk size for a typical event.
    }
    \label{fig:chunk_speedup}
\end{figure}

Given the ability to predict the peak memory usage for each event, we can adapt the pixel chunk size to maximize usage of the available memory and reduce the computation time. The speed-up factor from tuning this chunk size relative to using a chunk size of 1 pixel for a typical event is shown in Fig.~\ref{fig:chunk_speedup}. This speed-up is not linear because of the finite number of computing units for the given GPU, which limits the number of parallel operations. In general, this technique achieves a speed-up of around a factor of 5 for most of the events, and it enables efficient use of GPU memory.

\section{Parameter fitting}
\label{sec:param-fitting}

In the following, we will apply the differentiable simulator described above to the task of calibration. In our context, this means tuning parameters of our simulator to match some given dataset, which can be either simulated or from a real experimental setup.
Let $f(\chi, \theta)$ represent our differentiable simulator, with input particle segments $\chi$ and parameters $\theta$. Our focus is to optimize the parameters $\theta$.

For the optimization of $\theta$, we propose an ``analysis-by-synthesis'' approach:
\begin{enumerate}
    \item Choose initial values for the parameters, denoted as $\theta_0$;
    \item Run the forward simulation with these parameters, $f(\chi, \theta_0)$;
    \item Compare the simulation output with target data $F_\text{target}$, using a \textit{loss function}, $\mathcal{L}(f(\chi, \theta_0), F_\text{target})$;
    \item Update parameter values $\theta_0 \rightarrow \theta_i$ to minimize the loss, and repeat from step $2$ starting from parameters $\theta_i$ and forward simulation $f(\chi, \theta_i)$. The differentiable simulator enables a gradient-based update rule for $\theta_i$.
\end{enumerate}

The ultimate goal for this parameter fitting approach is an application to real, measured data. 
For measured data, we cannot access the true particle energy depositions, and the fitting procedure must include inference of particle segments $\chi$ from the data. If the simulation $f$ describes the measured data well, we may use an analysis-by-synthesis approach to find parameters $\theta$ and particle segments $\chi$ which best describe a measured $F_{\text{target}}$. We cannot know the ``real'' parameter values $\theta_\text{target}$ or particle segments $\chi_{\text{target}}$ which produced the measured $F_{\text{target}}$, but the procedure will produce simulation outputs which closely mimic the data, and fitted parameter values and segments may provide insight on experimental settings.

To demonstrate the capability of our differentiable simulator in optimizing model parameters, we focus on a controlled case where $F_{\text{target}}$ is generated using our simulator. Simulated targets $F_\text{target}$ are constructed as $f(\chi, \theta_\text{target})$, where particle segments $\chi$ are known and may be used directly in the fit. Similarly, $\theta_{\text{target}}$ are known parameter values producing $F_\text{target}$.
We can therefore benchmark the performance of our optimization by comparing fitted parameters $\theta$ with these known values of $\theta_\text{target}$. Successful fits recover fitted values $\theta = \theta_\text{target}$. This procedure is therefore known as a closure test.

Because our simulator, $f$, is differentiable, we can update the parameter values (step 4) using gradient-based optimization algorithms.
Defining a differentiable loss function $\mathcal{L}$, we are able to efficiently calculate $\nabla_{\theta} \mathcal{L}(f(\chi, \theta), F_\text{target})$.
For, e.g., gradient descent, the parameter update then takes takes the form
\begin{equation}
    \theta_{i+1} = \theta_{i} - \eta \cdot \nabla_{\theta} \mathcal{L}(f(\chi, \theta_{i}), F_\text{target})
\end{equation}
for iteration step $i$, where $\eta$ is a learning rate which controls the size of the update. In practice, more sophisticated update rules (e.g. Adam~\cite{Adam}) may provide better convergence.

\begin{figure*}
  \centering
  \includegraphics[width=0.7\linewidth]{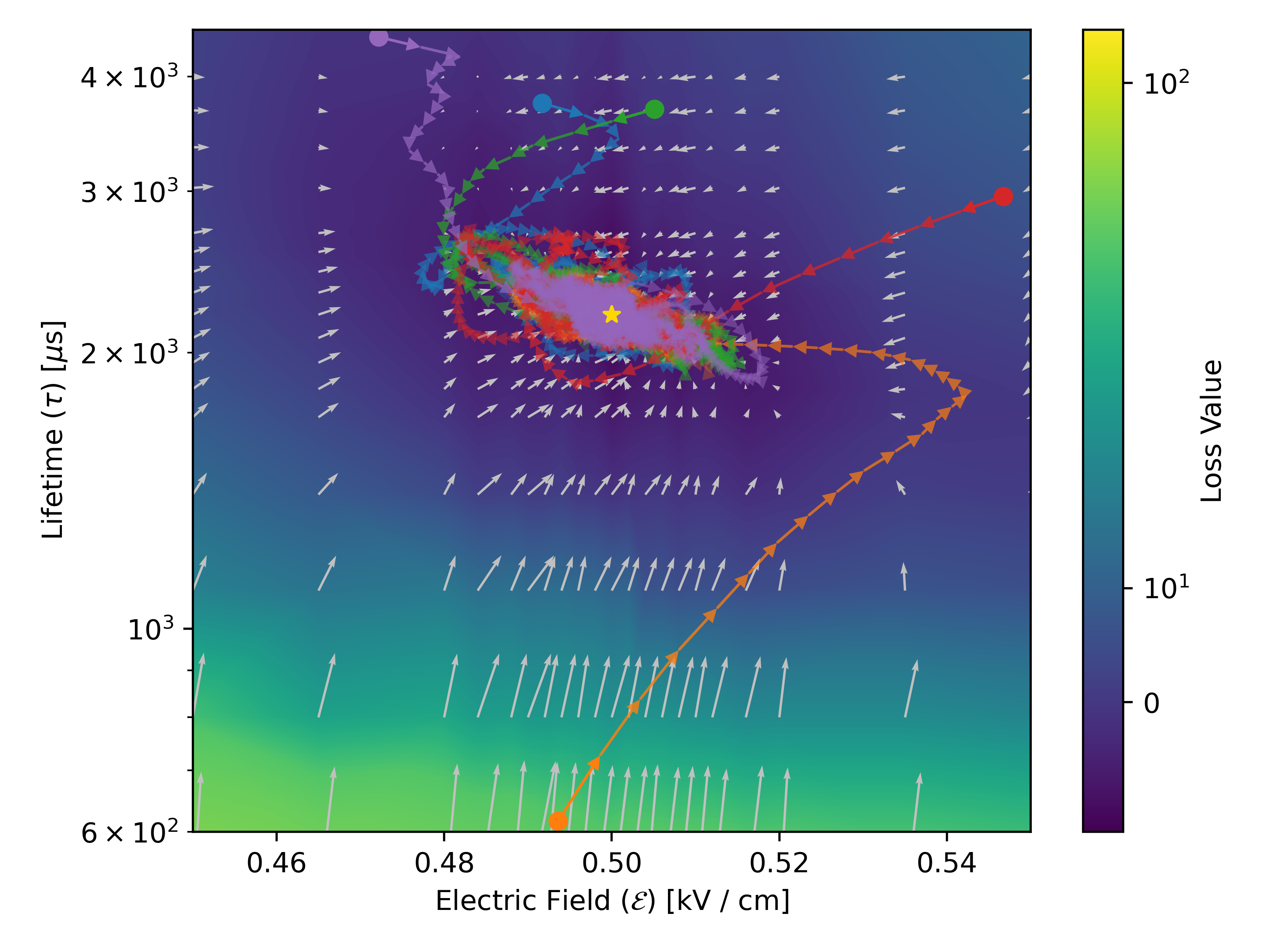}
  \caption{Loss landscape in a 2D parameter space of electric field $\mathcal{E}$ and lifetime $\tau$, averaged across batches. The gold star labels the target parameter values. The negative gradient, shown in the white arrows, points towards the loss landscape minimum at the target. Five example fit trajectories, starting from a variety of different initial points (filled circles) are shown respectively in different colored lines. All fits converge to the target parameter values (the gold star).
  }
  \label{fig:loss-landscape}
\end{figure*}

This fitting procedure is illustrated in Fig.~\ref{fig:loss-landscape}.
For visual simplicity, we restrict the illustration to the 2D parameter phase space of electric field $\mathcal{E}$ and lifetime $\tau$.
The gold star indicates a set of target parameter values in this 2D phase space.
The background color shows the loss landscape, where the loss function is evaluated with respect to the simulated target for a simulation at each sampled point.
The white arrows across the loss landscape indicate the negative gradients, corresponding to the direction of a gradient descent step with our differentiable simulator. These arrows point towards the minimum of the loss at the target parameter point.
The progression of 5 fits in the parameter space, labeled by color, with different initial parameter values (filled circles) is overlaid on the loss landscape.
The colored arrows indicate the parameter update per iteration step.
All 5 fits converge to the target parameter values.

For the demonstration of parameter fitting with our differentiable simulator, we select 6 commonly considered detector model parameters which are listed in Table~\ref{table:param_ranges} and highlighted in Fig.~\ref{fig:simulation-flow}.
A description of these parameters can be found in Sec.~\ref{sec:sim-description}.
The nominal values of the parameters are set to the default parameter settings of our snapshot of \texttt{larnd-sim}.
We define a relevant physical range for each parameter which covers a range of possible parameter values from a variety of measurements. These ranges are typically larger than the uncertainties on a single, specific measurement.
Among these 6 parameters, the Birks model parameters, $A_B$ and $k_B$, and the diffusion parameters, $D_L$ and $D_T$, are generic model parameters for the general LArTPC system, and therefore are not expected to vary significantly between experiments. In contrast,the electric field, $\mathcal{E}$, and electron lifetime, $\tau$, are operational LArTPC model parameters that depend on particular experimental processes and settings. This means that the range of $\tau$, for example, is relatively wide, as it depends on how well the liquid argon is purified, which has minimal external constraint.

\begin{table}[h!]
\centering
\resizebox{0.48\textwidth}{!}{
\begin{tabular}{| l | l | l |}
\hline
\textbf{Parameter [Units]} & \textbf{Nominal Value} & \textbf{Range}  \\ 
\hline \hline
$A_{B}$ & 0.8~\cite{Birks} & [0.78, 0.88]~\cite{Box}\\
$k_{B}$ [$kV.g/cm^3/MeV$] & 0.0486~\cite{Birks} & [0.04, 0.07]~\cite{Box}\\ 
$\mathcal{E}$ [$kV/cm$] & 0.5 & [0.45, 0.55]~\cite{MicroBooNE:2019koz} \\ 
$\tau$ [$\mu s$] & 2200 &  [500, 5000]\\ 
$D_{L}$ [$cm^{2}/\mu s$] & $4\times 10^{-6}$ & [$2\times 10^{-6}$, $9\times 10^{-6}$]~\cite{vdrift-param, MicroBooNE:2021icu}\\
$D_{T}$ [$cm^{2}/\mu s$] & $8.8\times 10^{-6}$ & [$4\times 10^{-6}$, $14\times 10^{-6}$]~\cite{vdrift-param, MicroBooNE:2021icu} \\
\hline
\end{tabular}}
\caption{6 fit parameters with corresponding nominal values and defined physical ranges used for this study.}
\label{table:param_ranges}
\end{table}

With the differentiable simulator, all six parameters of interest are optimized simultaneously.
This has not been achieved in conventional calibration due to the limitations of the associated methodology. We briefly describe the conventional approach for setting each of these parameters.

In conventional calibration, the Birks model parameters, $A_B$ and $k_B$, are jointly fit using a control sample. This control sample allows for estimation of $\frac{dE}{dx}$ as a prior for the fit, and the electric field $\mathcal{E}$ is set to a fixed value (Eq.~\ref{eq:Birks}).

The value of the electric field $\mathcal{E}$ is often calculated using the measured voltage on the cathode divided by the maximum detector drift distance (the distance from cathode to anode), assuming the cathode and anode are perfectly parallel planes.
Deviations from this assumption, as well as shrinkage of the detector under cryogenic temperatures during operation, can shift $\mathcal{E}$ away from the nominal value.
Furthermore, the actual electric field across the detector may not be perfectly uniform. However, this is often a sub-leading effect.

The electron lifetime $\tau$ is commonly measured with a muon control sample, which has relatively uniform $\frac{dE}{dx}$. In order to fit $\tau$, the recombination model parameters, $A_B$ and $k_B$, and the electric field, $\mathcal{E}$, are set to fixed values.

The measurement of the longitudinal diffusion coefficient $D_L$ often uses a control sample of muons due to their uniform detector signal.
Nominally $D_L$ affects the signal extension in terms of drift time (width of the readout waveform).
In order to extract $D_L$, we need to separate out the effects from drift velocity and drift time.
The electron lifetime and the electric field can both affect the shape of the readout waveform, and therefore need to be well understood. However, they typically are not explicitly treated in $D_L$ measurements. The transverse diffusion coefficient, $D_T$, is usually extracted based on $D_L$, assuming a fixed value of the electric field and a given electron mobility model $\mu(\mathcal{E}, T)$.

\begin{figure}
  \centering
  \includegraphics[width=\linewidth]{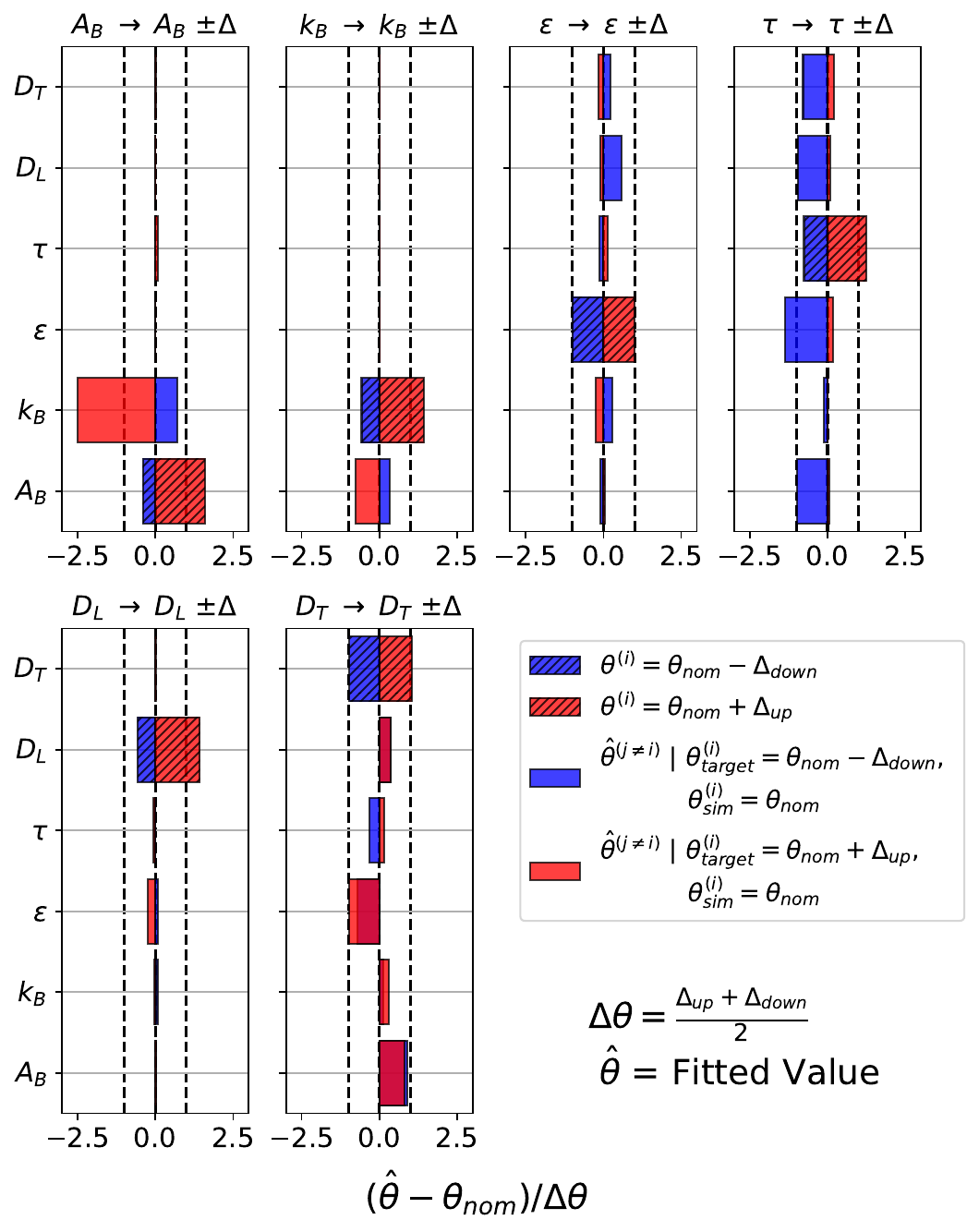}
  \caption{Demonstration of interdependence between parameters. Parameters with hatched fill correspond to those which are shifted to their lower (blue) or upper (red) range limits in the target simulation and fixed to their nominal values in the corresponding fit. The colored bars show the corresponding bias from the fitted result.
  }
  \label{fig:param-interdep}
\end{figure}

In these conventional approaches, the determination of particular parameters is often done assuming fixed values of the other parameters. In practice, different detector processes can effect the measured data in similar ways, meaning that this assumption may cause incorrect results.
This is demonstrated in Fig.~\ref{fig:param-interdep}, which shows that if a bias exists in a model parameter which is fixed in the calibration, other parameters of interest may converge to biased values. 
In the figure, nominal values of the parameters are denoted by $\theta_\text{nom}$.
$\Delta_\text{down}$ and $\Delta_\text{up}$ are the distances from $\theta_\text{nom}$ to the lower and upper boundaries of the range shown in Table~\ref{table:param_ranges}.
For each parameter respectively, $\Delta\theta$ is the average of $\Delta_\text{down}$ and $\Delta_\text{up}$.
Each panel shows the biases of the fitted parameter values in units of $\Delta\theta$ resulting from a shift of one selected parameter to its lower (blue) or upper (red) range in the fitting target while fixing that same parameter to its nominal value during the fit optimization. 
All other parameters are set to their nominal values in the target.
In several cases, fitted parameters deviate significantly from their target values. Therefore, optimizing all detector parameters simultaneously is important for avoiding calibration biases and achieving precision physics modeling.

\subsection{Loss function}
\label{subsec:loss}

Optimization of fitting parameters requires the construction of an appropriate loss function, $\mathcal{L}$. The properties of this loss function have a significant impact on the optimization. We describe several options and requirements in order to motivate our chosen approach.

The loss function must be differentiable with respect to the parameters of interest. It must be well-behaved when comparing two arbitrary simulation outputs. For the closure test, with no electronics noise, the loss function should give the global minimum when the fitted value $\hat{\theta} = \theta_\text{target}$, i.e. when two simulation outputs are identical. Similarly, loss values should increase as simulation outputs become more different, which should correspond with $\theta$ deviating from $\theta_{\text{target}}$.

The output of the simulation is the charge $q$, in ADC counts, read out on relevant pixels of coordinates $x$ and $y$ at time $t$. Pixel coordinates and readout times are discrete; it is possible to use these discrete bins to voxelize the entire LArTPC readout space in 3D.
However, considering the number of pixels and the readout sampling frequency, a full resolution binning of the LArTPC readout would correspond to $O(10^9)$ floating point numbers.
Building a loss function, such as a mean-squared error, on these full resolution voxels involves comparing two such dense tensors, which is a very costly operation.

Typically, only a small fraction of the 3D space has recorded charge in the readout.
One can, for instance, select rectangular regions of voxel space that only contain the active 3D voxels and their neighbors.
This is less expensive, but still may be expected to contain empty voxels. Furthermore, the selected voxels for comparison must contain both the target readout and the output of each optimization step, and may still require a large number of voxels for sufficiently populated events or to contain sufficiently separated particle segments. 

It is more efficient to carry out a ``sparse'' comparison by only calculating a loss where charge readout occurs.
Instead of a rectangular voxel grid, we can frame the readout as a sequence of points, $\{(x_i, y_i, t_i, q_i)\}_{i=1}^n$, where $n$ is the total number of the active readout points.
Pixels and times with null readout are excluded from the loss calculation.
$x$, $y$, $z$, and $q$ are treated as continuous variables in this construction -- voxelization never explicitly enters the calculation.

A natural way to construct the loss function, $\mathcal{L}(\{(x, y, t, q)\}(\theta), \{(x_\text{target}, y_\text{target}, t_\text{target}, q_\text{target})\}(\theta_\text{target}))$, is to match readout points between the output of an iteration and the target. Compared to approaches that operate on the entire set or distribution of points in aggregate, this approach has the advantage of high granularity and sensitivity to changes in the readout. However, the point matching faces two major challenges: (1) $\{(x_\text{target}, y_\text{target}, t_\text{target}, q_\text{target})\}$ and $\{(x, y, t, q)\}$ may not have the same number of points. (2) There may not be a direct association between points $(x_\text{target}, y_\text{target}, t_\text{target}, q_\text{target})_i$ and $(x, y, t, q)_j$, even though they are simulated from the same set of particle segments.

In time series analysis, a dynamic time warping discrepancy (DTW)~\cite{DTW} is used for a very related problem. 
Due to variations in speed during the sequences, direct comparisons between points at particular times may miss shared patterns between sequences -- e.g. two identical sequences offset by 1 second may have large differences at a given set of times.
DTW addresses this challenge by computing an optimal alignment between two time series, under a chosen distance metric and a set of algorithmic constraints.
The two sequences can have different lengths, and multiple points from one sequence can be matched to a single point from the other sequence.
It is therefore very well suited for our task, and we adopt it as our loss function for the results presented here.

DTW requires a choice of ordering for the input sequences of points. For our loss function, we use the default ordering of our simulator outputs, which is a per-event ordering based primarily on pixel $x$ coordinates, with $y$ ($t$) used to break ties when $x$ ($y$) coordinates are shared between points.
This ordering largely keeps the structure of the input segments and corresponds well to the structure of the detector readout.

This is not a unique choice of ordering. As the readout planes are in $(x,y)$, instead choosing $y$ as the primary axis can be expected to yield similar results.
Although $t$ is a special axis in LArTPCs, an ordering with time as the primary axis would also retain the particle segment structure, and therefore should produce compatible fitting results.
Other options, such as a non-hierarchical combination of $x$, $y$, and $t$ coordinates, are also possible, but we expect little improvement relative to the current approach.

Once the points are ordered, there is a further choice of what features to compare between the two sequences. In principle it is possible to compare multi-dimensional features using the DTW loss. However, as geometric information is encoded into the ordering of sequences, and the primary impact of parameter variation for our chosen parameters is on the pattern of measured charge, it is sufficient to use only ADC counts. The inputs to the DTW loss function are therefore sequences of ADC counts, hierarchically ordered by the corresponding values of $x$, $y$, and $z$.

Dynamic time warping is not nicely differentiable by default. We therefore employ Soft-DTW~\cite{SoftDTW}, a smoothed version of DTW,
using the implementation from Ref.~\cite{SoftDTWCUDA-0, SoftDTWCUDA-1}.
More concretely, Soft-DTW uses a special ``$\min$'' operator with a smoothing parameter $\gamma \geq 0$. As $\gamma$ approaches $0$, this operator converges to the unsmoothed $\min$.
We set $\gamma = 1$ in our application. 
An absolute difference is used as the DTW distance metric in this work.

\subsection{Fitting considerations}
\label{subsec:procedure}

For this demonstration, we fit 6 parameters using our differentiable simulator: the Birks model parameters $A_{B}$, $k_{B}$; electric field $\mathcal{E}$; lifetime $\tau$; and longitudinal and transverse diffusion coefficients $D_{L}$ and $D_{T}$. 
The physical values of each of these parameters range across several orders of magnitude relative to each other (Table~\ref{table:param_ranges}). This can be challenging for the optimization, as it results in a large range of magnitudes for the corresponding gradients with respect to each parameter.
To balance relative scales among the parameters, we normalize the parameters using their nominal values, $\bar{\theta} = \frac{\theta}{\theta_\text{nom}}$. 
On each fit iteration, gradient calculation is done with respect to the normalized parameters, resulting in more balanced updates among the parameter set. We therefore avoid needing to carefully tune individual parameter learning rates for simultaneous parameter fits.
The physical parameter value is needed for the simulation itself. We therefore undo the normalization before running the forward model simulation at each iteration, $f(\chi, \bar{\theta}_{i}\cdot \theta_{\text{nom}})$.

We use Adam~\cite{Adam} as our gradient-based optimizer. With the normalization, we can set a single learning rate, $5\times 10^{-2}$, for all parameters.
In addition, an exponential learning rate scheduling with a decay rate of 0.95 is adopted to aid stability of convergence, and the learning rate is updated using the scheduler after each epoch (one pass through the full dataset). We clip the norm of the vector of all normalized parameter gradients at 1 to avoid large iteration steps.

Each fitting iteration is performed using a mini-batch of simulated data, which is a subset of the entire input dataset. The mini-batches are distinct subsets that in total cover all of the input data, and the division of data across mini-batches remains the same across every epoch in a given fit. To construct the mini-batches, the full dataset of particles, with their constituent particle segments in order, are randomly shuffled. We do not mix the particle segments so that we maintain the impact of induced current from neighboring charges (see \textbf{Current accumulation} in Sec.~\ref{sec:sim-description}).
As discussed in Sec.~\ref{subsec:computation}, total particle segment length $ds$ provides a good measure of mini-batch computational time. In mini-batch construction, segments from the shuffled particles are sequentially added to each batch until the total length of segments in the mini-batch reaches a chosen value of $ds$. For this study, we choose to make mini-batches of $ds=\SI{100}{\cm}$.

Losses are computed event-by-event and then averaged across events in each mini-batch. Gradients are calculated based on this mini-batch loss, and parameters are updated correspondingly. Because each iteration is done with a single mini-batch, the parameters of interest can be updated multiple times per epoch, depending on the total number of mini-batches. This aids computational efficiency of the optimization, as the amount of data processing required per iteration is greatly reduced. 

\begin{figure}
\centering
 \includegraphics[width=\linewidth]{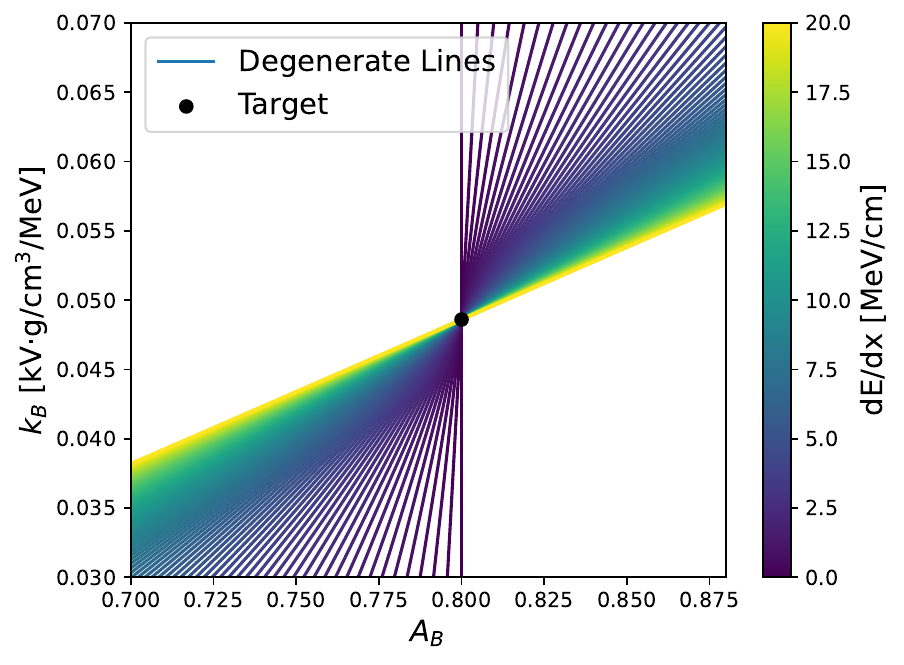}
 \caption{Degeneracy between the Birks model parameters $A_B$ and $k_B$ if the energy deposition per unit length $\frac{dE}{dx}$ is a single value. With a range of $\frac{dE}{dx}$, the degeneracy breaks.
 }
 \label{fig:Ab-kb-degen}
\end{figure}

Because parameter calibration using our differentiable simulator relies on the correspondence between parameters $\theta_{\text{target}}$ and the resulting detector readout, we pay special attention to potential degeneracies among parameters -- cases where the detector readout is the same, even if parameter values are different. We focus here on degeneracies within the physical models. Degeneracy can also be introduced during other pieces of the simulation (e.g. due to electronics noise).
We highlight two potential model degeneracies:
\begin{enumerate}
    \item For the Birks model in isolation, there is a degeneracy between $\mathcal{E}$ and $k_{B}$ due to the term $k_{B} / \mathcal{E}$ (Eq.~\ref{eq:Birks}). This means that simultaneous fits of $k_{B}$ and $\mathcal{E}$ will not necessarily converge to their target values -- as long as the ratio $k_{B} / \mathcal{E}$ converges to its target, the individual parameter values do not matter. In the context of the broader simulation, however, this degeneracy is broken due to the relation $v_{\text{drift}} = \mu(\mathcal{E}) \cdot \mathcal{E}$, which results in impacts from the electric field in parts of the simulation outside of the Birks model. Note that a similar degeneracy exists in the Box model, which is an alternative recombination model. In conventional calibration methods, $\mathcal{E}$ is usually considered as ``well-measured'' and $\mathcal{E}$ is therefore not fit, breaking this degeneracy. 

    \item Another degeneracy exists in the Birks model. 
    Assume that the readout can be described with an electron survival rate $\alpha_{\text{recomb}}^{*}$ from the recombination. For a given particle segment $\frac{dE}{dx}$ and fixed $\mathcal{E}$ and $\rho$, there is a set of values of $A_{B}$ and $k_{B}$ which will produce the same value of $\alpha_{\text{recomb}}^{*}$. This set is given by the linear relation
    \begin{equation}
         A_{B} = \alpha_{\text{recomb}}^{*} \cdot \Big(1 + \frac{k_{B}}{\mathcal{E}\cdot \rho}\frac{dE}{dx}\Big).
    \end{equation}
    This degeneracy only holds for a single, fixed $\frac{dE}{dx}$. As illustrated in Fig.~\ref{fig:Ab-kb-degen}, changing $\frac{dE}{dx}$ changes the slope of this degenerate line in $A_{B}$ and $k_{B}$. The only parameter values that result in $\alpha_{\text{recomb}}^{*}$ for all $\frac{dE}{dx}$ will be the ``observed'' values, the $A_{B}$, $k_{B}$ point where all lines cross. An input dataset with a finite spread in $\frac{dE}{dx}$ is therefore required to break the degeneracy in $A_{B}$ and $k_{B}$. Further, the impact of this degeneracy will be reduced by fitting on a dataset with large variation in $\frac{dE}{dx}$. 
\end{enumerate}

The simulated closure test presented here is meant as a first demonstration of calibration fits using our differentiable simulator. We provide a starting point for two paths towards extending this procedure in support of the ultimate goal of application to real, measured data.

With real data, we do not have access to true particle segments or the corresponding true energy depositions. However, suitable input data for an optimization using our framework may be particles which typically produce track-like topology in LArTPCs, within an energy range of interest. With this topology, it possible to model particle segments by fitting lines to particle tracks and breaking them into segments. The $\frac{dE}{dx}$ of these segments can be inferred using the readout charge and the segment length. The performance of the parameter optimization will depend on the quality of the input segment estimation, and the segment estimation may be iteratively refined along with the parameters.

The studies presented here are done in the absence of electronic noise. This is because electronic noise introduces stochasticity in the correspondence between simulation parameters and the corresponding outputs, meaning that, even with the same input particle segments $\chi$ and parameter values $\theta$, we will get a different readout every time we run the forward simulation $f(\chi, \theta)$. Changes in the value of the loss therefore are not solely from changes in the parameters of interest, and target parameter values may not achieve a global minimum for a given sample of the electronics noise. 

In a real experimental setup, electronics noise cannot be easily disentangled from the measured data. It is therefore important to define a procedure for optimizing detector models using noisy data and to characterize the impact of the noise relative to the impact of parameter variation.

In data, a noisy target, $F_\text{target}$, is unavoidable. In simulation, however, we have full control over the noise model. We therefore suggest using the simulated forward model, $f(\chi, \theta)$, with the noise turned off for a fitting procedure with a noisy target. This reduces the overall stochasticity, and may aid the fit.

\begin{figure}
    \centering
    \subfloat[No electronics noise]{
         \includegraphics[width=0.95\linewidth]{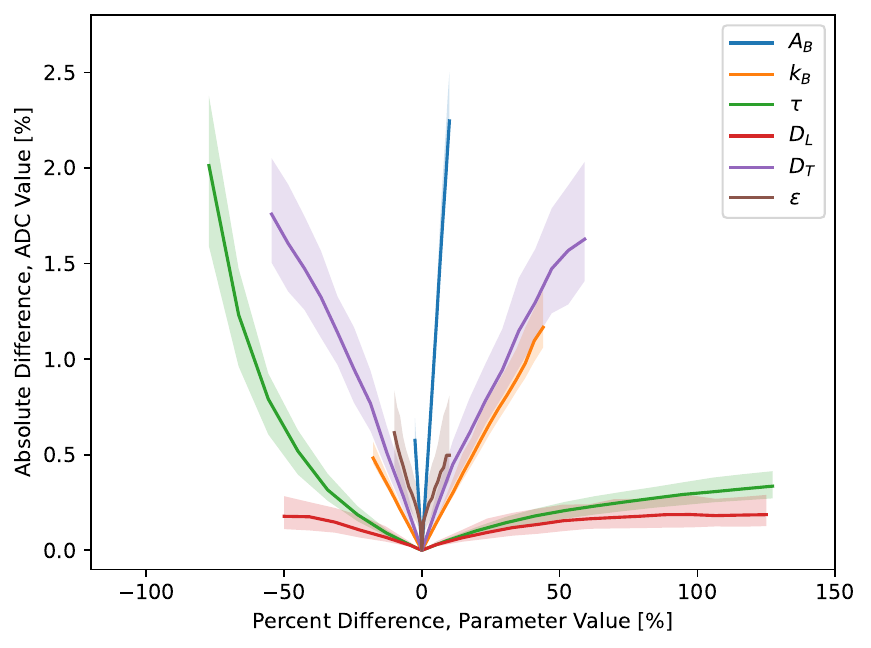}
         \label{fig:spider-plot-no-noise}}

    \subfloat[Electronics noise in targets]{
         \includegraphics[width=0.95\linewidth]{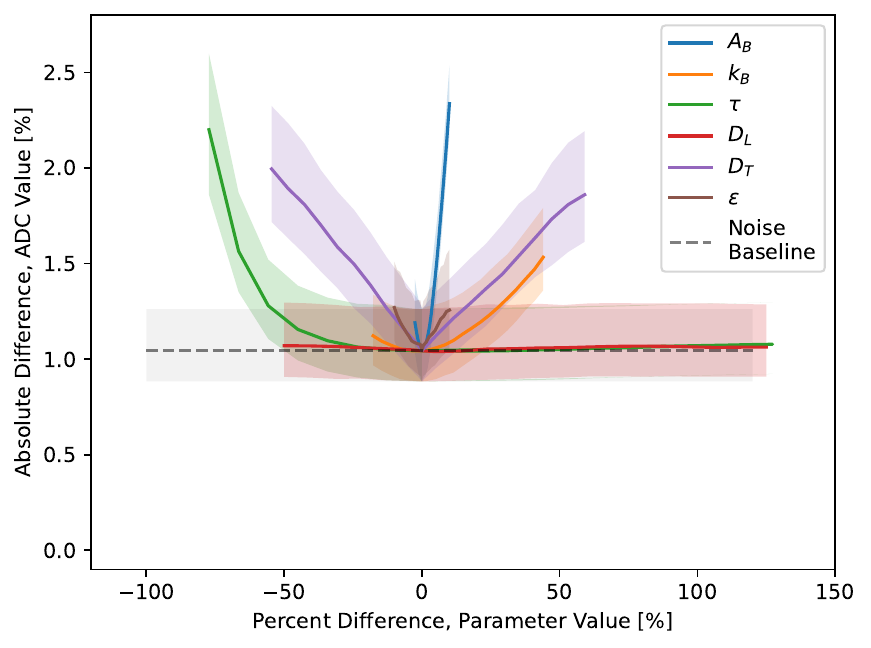}
         \label{fig:spider-plot-with-noise}}
         
    \caption{The lines show the absolute impact of each parameter on simulation output with respect to changes in the parameter value. The width of the band indicates variations of the impact across mini-batches. (a) The parameter impact is compared to a noiseless target. For all the parameters, the impact shows sharply defined minima at the nominal parameter values. (b) The parameter impact is compared to noisy targets. Minima are less well defined, and some parameter variation is indistinguishable from noise.
    }
    \label{fig:spider-plot}
\end{figure}

In Fig.~\ref{fig:spider-plot}, we present studies on the impact of electronics noise with respect to parameter variation using the default noise magnitude included in \texttt{larnd-sim}. Both sub-figures show the estimated impact of variations of individual parameters on the simulation output. Fig.~\ref{fig:spider-plot-no-noise} represents the case without electronic noise, while Fig.~\ref{fig:spider-plot-with-noise} shows the case with electronic noise in the target. For both cases, we do not include electronic noise in the simulations with varied parameter values, corresponding to the configurations of the closure test and our suggested procedure for fitting with noise respectively.

In Fig.~\ref{fig:spider-plot-no-noise}, a set of simulations is run for each parameter in which the parameter value is swept across the range defined in Table~\ref{table:param_ranges}, while all other parameters are kept at their nominal values. 
These parameter variations are shown as percentages of their corresponding nominal values. A target simulation is run with all parameters at their nominal values. For Fig.~\ref{fig:spider-plot-no-noise}, this target does not include electronic noise.
Each of these simulations is run across all mini-batches, and mini-batches are the same across all simulations. Loss values are calculated per mini-batch using the unsmoothed DTW described in Sec.~\ref{subsec:loss}. These losses are translated into an absolute percentage difference relative to the target simulation using a reference simulation which provides a constant scaling factor per mini-batch. The solid lines show the median value of this parameter impact across all mini-batches for each parameter, and the corresponding band shows the interquartile range across mini-batches, expressing batch-to-batch variation.
All parameters have sharp and clear minima at their nominal values, where they identically match with the target. For the majority of their respective ranges, $D_L$ and $\tau$ have relatively small impact, and may be expected to have the least sensitivity in the fit. The impact of $\tau$ is very asymmetric with respect to the nominal value.

Fig.~\ref{fig:spider-plot-with-noise} shows the same impact of parameter variation on the loss with respect to targets that include electronic noises. For each mini-batch, we make 10 corresponding targets with different samples of electronic noise. For every target, we then produce the same plot as in the no noise case. Translation of loss values to percentage impact is done using the same reference value as in the no noise case so that Fig.~\ref{fig:spider-plot-no-noise} and Fig.~\ref{fig:spider-plot-with-noise} can be compared directly. Similarly, the same mini-batches are used for both cases. The solid lines and band limits shown in Fig.~\ref{fig:spider-plot-with-noise} are the average of the corresponding median lines and their band limits across the 10 noise samples. The batch-to-batch variation expressed by the bands includes the impact of batch-to-batch variation in the noise. 

For reference, a noise baseline is included in Fig.~\ref{fig:spider-plot-with-noise}, which expresses the difference in readout values between simulations with and without noise with no parameter variation. The line is included across a range of parameters to guide the eye, but it is only calculated once, at the nominal parameter values. Calculation of this baseline is done across 10 noise samples identically as in the varied parameter case.
The impacts for a range of variations in $D_L$ and $\tau$ have flat shapes and poorly defined minima, showing a high degree of overlap with this noise baseline. This suggests that the impact of these parameter variations is not well distinguishable from the impact of noise. 

Electronic noise depends on the simulated detector configuration and may be reduced with post-processing steps. Mini-batch size and fitting data sample will also have an effect on the impact of noise. Closure test studies, such as the one presented in this work, demonstrate what is possible in an ideal scenario, and may provide a good baseline for analyzing loss of resolution due to effects such as electronic noise. We therefore suggest such studies as the starting point for particular applications.

\section{Samples used}
\label{sec:data}

\begin{figure}
  \centering
  \includegraphics[width=\linewidth]{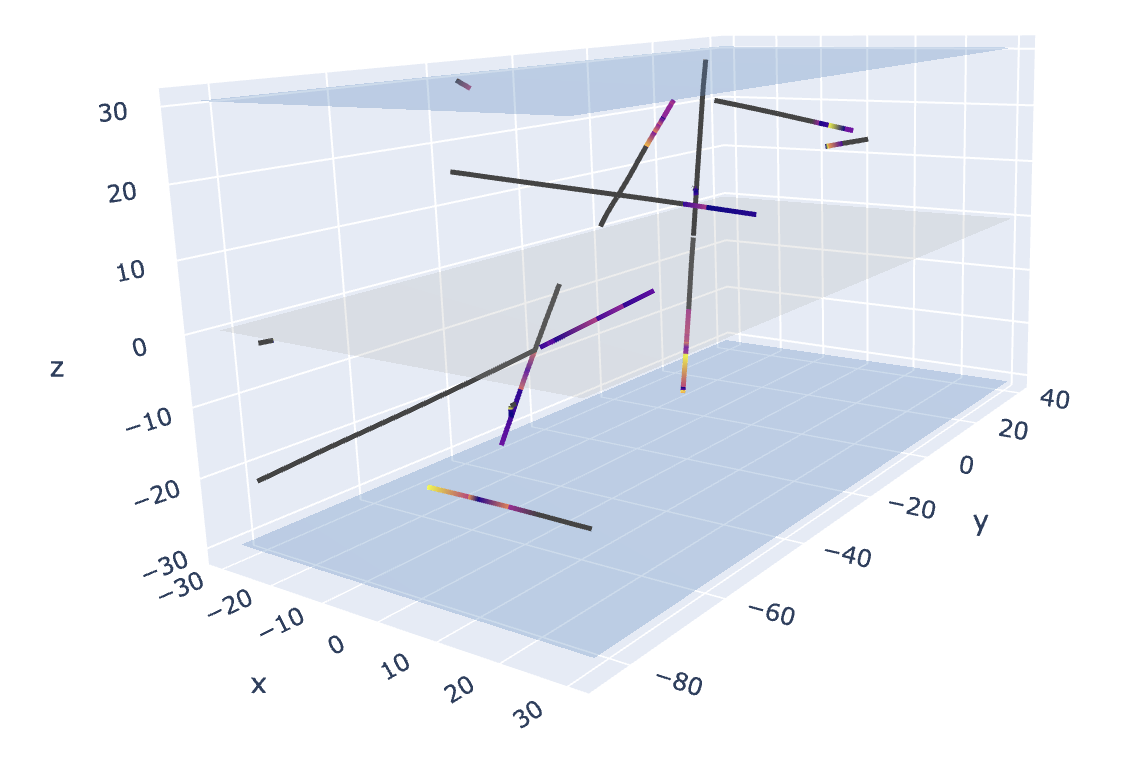}
  \caption{ A display of an event containing ten \SI{1}{\GeV} muons in the detector. The track color indicates the $\frac{dE}{dx}$ along the muon trajectories. The grey plane in the middle is the cathode, and the two blue planes are the anodes of the two TPCs. 
  }
  \label{fig:event_display}
\end{figure}

Typically, control samples, or samples with well known properties, are used to understand detector modeling.
Muons are a good candidate for this purpose.
As minimum ionizing particles, their energy depositions in liquid argon ($\frac{dE}{dx}$) are narrowly peaked around \SI{1.6}{\MeV\per\cm}. There is, however, sufficient enough spread in $\frac{dE}{dx}$ to break the degeneracy discussed in Sec.~\ref{subsec:procedure}.
In an experimental setup, samples of cosmic muons are also relatively easy to obtain.
To mimic a muon control sample, we simulate 100 events with about 10 muons per each event, injected into the LArTPC volume from the detector border. Simulation is done using DLPGenerator~\cite{DLPGenerator} and edep-sim~\cite{edep-sim}.
All muon kinetic energies are set to be \SI{1}{\GeV}, and muon injection angles are sampled randomly from an isotropic distribution.
Fig.~\ref{fig:event_display} shows an example of one event, where the lines indicate the muon trajectories (particle tracks). 
The particle tracks are composed of segments. 
Each segment is modeled as a straight line of constant $\frac{dE}{dx}$, and the length of the segment can vary based on the rate of change of the energy deposition. 
See Sec.~\ref{sec:sim-description} for the description of how the segments are used in the simulation.

For detector calibration, $O(1000)$ muons is a very small sample compared to what is commonly used.
It would have a negligible impact on the regular data taking to acquire such a small control sample, even if this acquisition was done frequently. However, Sec.~\ref{sec:results} shows that this seemingly small muon sample is sufficient for calibrating detector model parameters.
This simulated muon sample is therefore used as the default sample to benchmark the performance of the differentiable simulation.

We have also produced a mixed particle sample that is composed of muons, pions, and protons with kinetic energies uniformly sampled from \SIrange[]{0.1}{2}{\GeV}.
This is to test whether we can relax the requirement of muons as calibration data and to see how robust the parameter optimization is to different particle types and energies. 

To ensure a stable convergence with a reasonable computational allocation, we apply four selection cuts to the input data. These cuts are not optimized to maximize data usage, and may be stricter than necessary.

We remove tracks that are almost colinear to the drift axis $z$ by requiring abs($\cos{\phi}$)$ < 0.966$, where $\phi$ is the angle between the track and $z$. This cut helps to control memory usage, as discussed in Sec.~\ref{subsec:computation}.

We only include segments whose centers are within \SI{15}{\cm} of drift distance ($z$) to the anodes. 
The maximum drift distance for our detector configuration is \SI{30}{\cm} (Sec.~\ref{sec:sim-description}), so this corresponds to excluding segments that lie in the region near the cathode. 
We also remove tracks that have any segment whose center is within \SI{2}{\cm} of either of the two anodes.

\begin{figure}
  \centering
  \includegraphics[width=\linewidth]{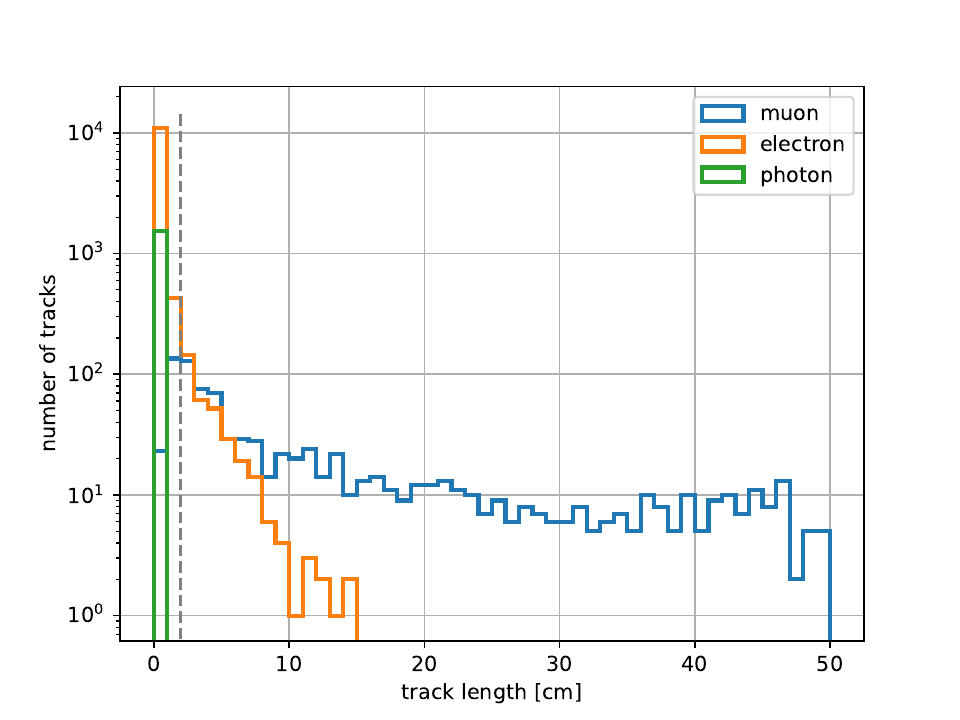}
  \caption{Particle trajectory length (sum of particle segment length $ds$) for simulated muons, electrons, and photons in the input muon sample. Cutting at $ds=\SI{2}{\cm}$ removes a significant fraction of electrons and photons. 
  }
  \label{fig:track_len}
\end{figure}

Muon interaction with the detector volume can trigger radiation photons and produce delta rays and Michel electrons.
We require track length to be longer than \SI{2}{\cm} to eliminate most of the delta rays, Michel electrons, and radiation photons (see Fig.~\ref{fig:track_len}), which often are topologically associated and even attached to the muon tracks. This aids fitting performance, and prevents computation time from being dominated by many short, low energy particle segments.

It is possible to do a calibration fit without the above selections. However the selections help to constrain computational requirements and to improve the optimization landscape. 
As an example, we found that including particle segments with longer drift lengths introduced local minima, perhaps due to local degeneracies in readout hit topological patterns across multiple values of the diffusion parameters. 
These local minima decrease the robustness of the parameter optimization, as fits which encounter a local minimum become ``stuck'' and do not converge to the target parameter values.
Such fits are often distinguishable by a higher corresponding loss value and may be addressed by varying initial fit conditions and picking the result or results with the lowest loss. This latter process is robust, but requires additional computation and attention to the fit. We therefore adopt the above selection cuts, which avoid such issues.


\section{Results}
\label{sec:results}
We present results of simultaneously fitting six detector model parameters using the differentiable simulator: $A_B$, $k_B$, $\mathcal{E}$, $\tau$, $D_L$ and $D_T$, as listed in Table~\ref{table:param_ranges} and introduced in Sec.~\ref{sec:sim-description}. All fits are run on single NVIDIA Tesla A100 GPUs.

The results are from a simulation closure test, as described in Sec.~\ref{subsec:procedure}. 
To select parameter targets, we draw a uniform random value for each parameter from the ranges shown in Table~\ref{table:param_ranges}. Each target then corresponds to a single point sampled uniformly from the considered 6D phase space. All fits begin with parameters initialized at the nominal values shown in Table~\ref{table:param_ranges}. 
This setup mimics a realistic experimental procedure, as our best initial guess for each parameter is based on previous measurements.

\begin{figure}
     \centering
     \subfloat[Charge recombination parameter $A_{B}$]{
         \centering
         \includegraphics[width=0.49\columnwidth]{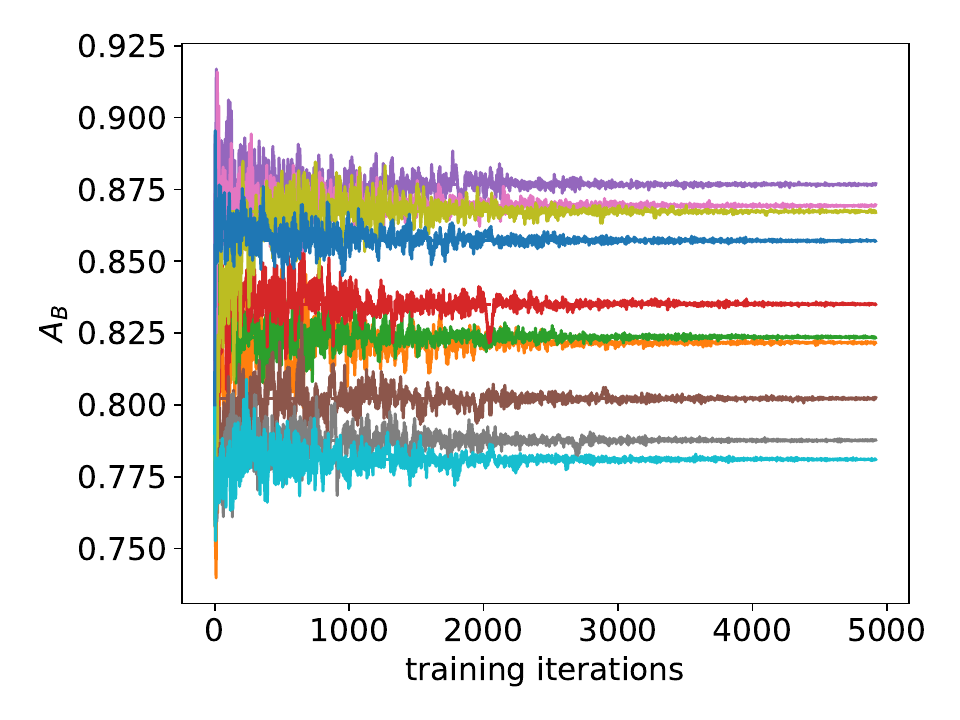}
         \label{fig:fit_result_Ab}}
     \subfloat[Charge recombination parameter $k_{B}$]{
         \centering
         \includegraphics[width=0.49\columnwidth]{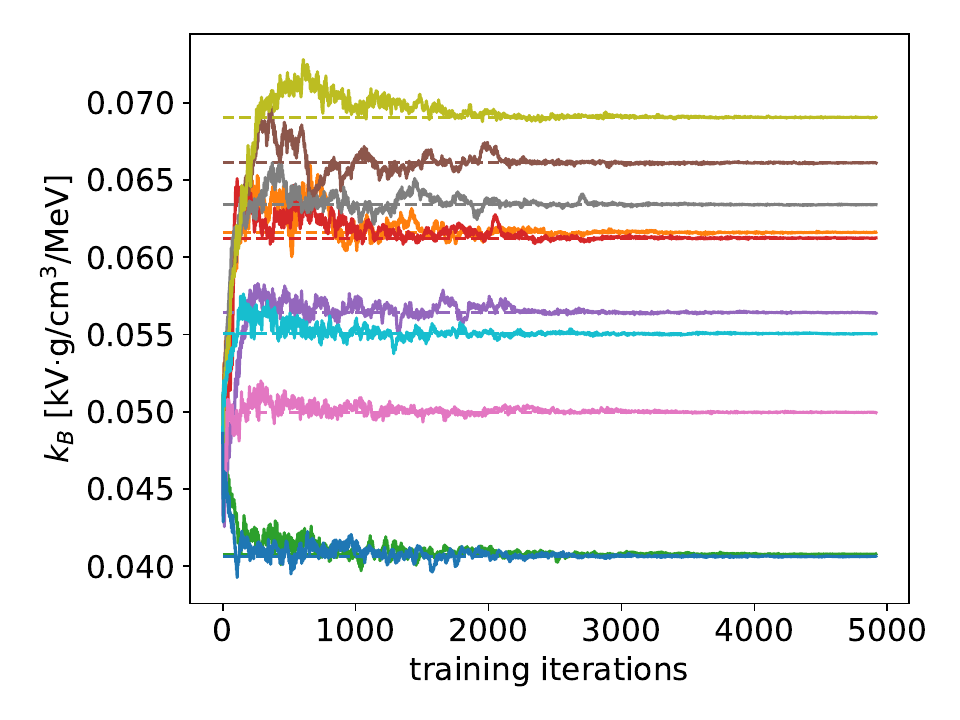}
         \label{fig:fit_result_kb}}
         
     \subfloat[Drift electric field $\mathcal{E}$]{
         \centering
         \includegraphics[width=0.49\columnwidth]{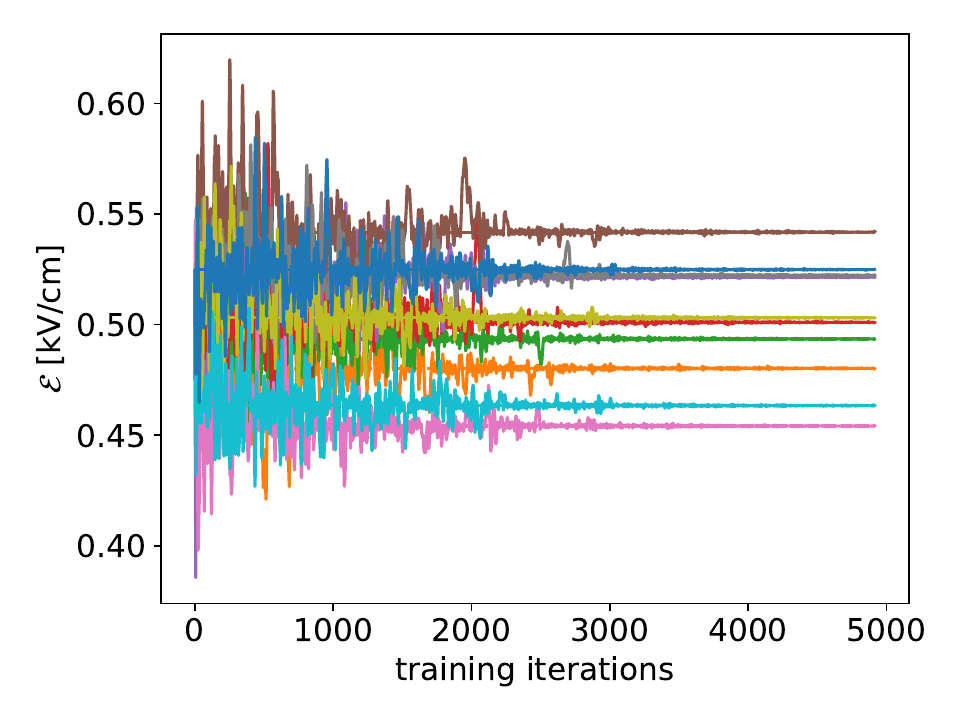}
         \label{fig:fit_result_eField}}
     \subfloat[Electron lifetime $\tau$]{
         \centering
         \includegraphics[width=0.49\columnwidth]{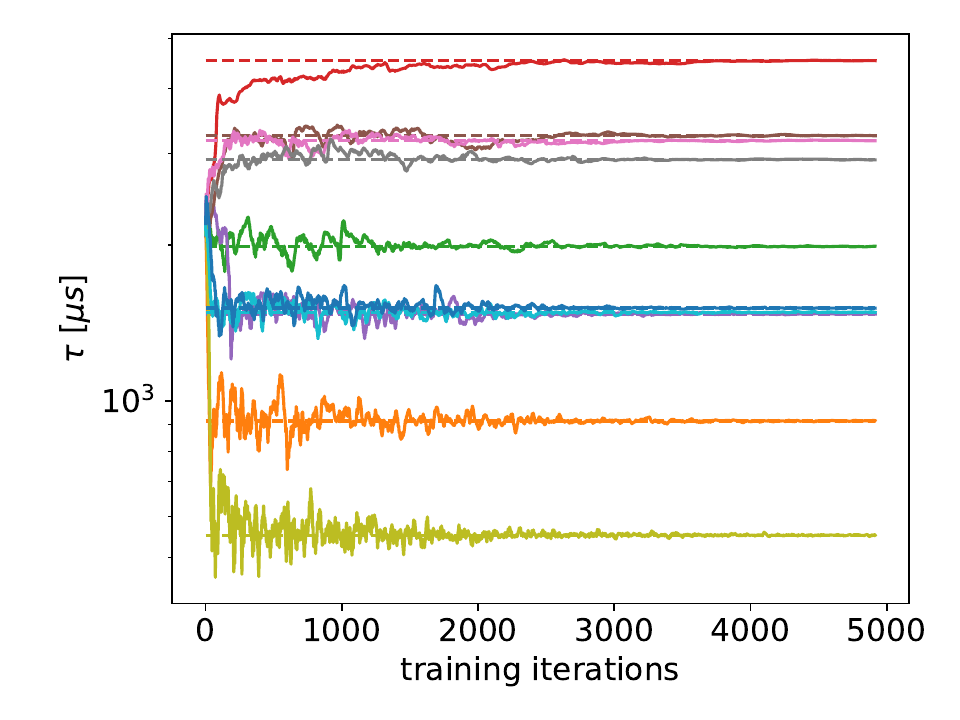}
         \label{fig:fit_result_lifetime}}
         
     \subfloat[Transversal diffusion parameter $D_{T}$]{
         \centering
         \includegraphics[width=0.49\columnwidth]{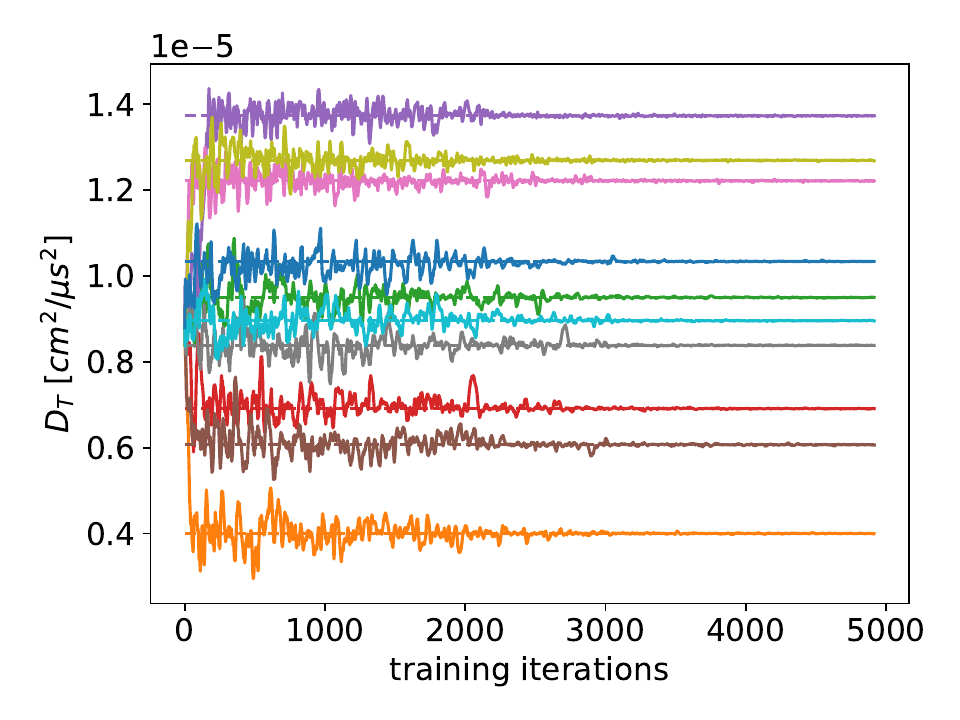}
         \label{fig:fit_result_tran_diff}}
     \subfloat[Longitudinal diffusion parameter $D_{L}$]{
         \centering
         \includegraphics[width=0.49\columnwidth]{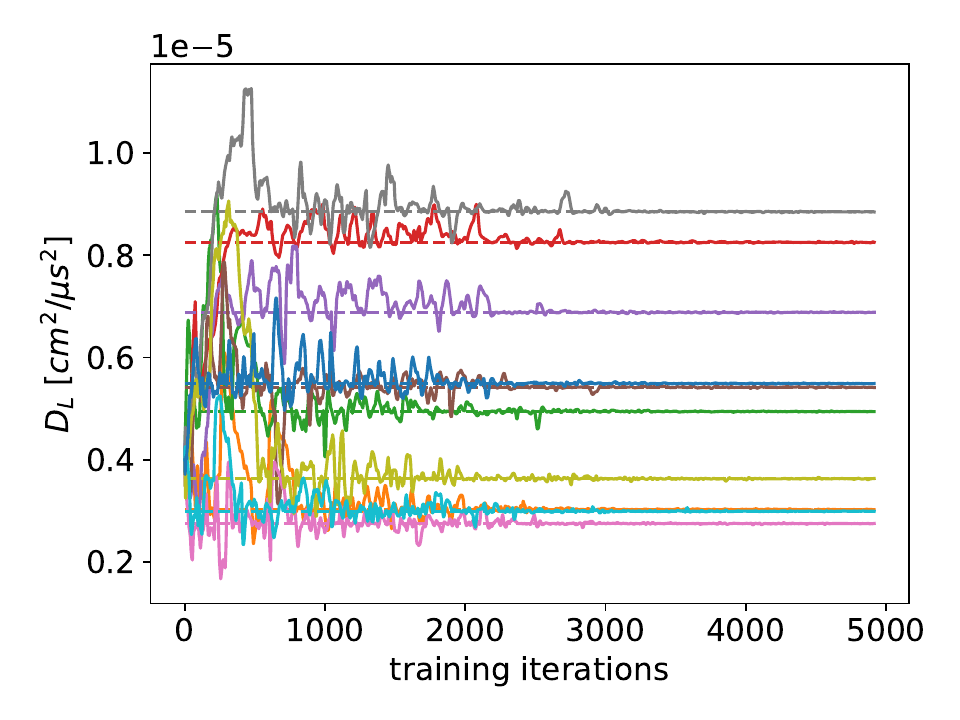}
         \label{fig:fit_result_long_diff}}
     \caption{Results from a simultaneous fit of six detector model parameters: $A_{B}$, $k_{B}$, $\mathcal{E}$, $\tau$, $D_{T}$, $D_{L}$ using the default muon sample. Each color indicates a sampled 6D target parameter point, and all fits start from the same initial parameter values, mimicking a realistic fitting scenario. The dashed lines label the target values of each parameter for each respective fit. The solid lines show the evolution of fitted parameter values with respect to the iteration number in the fit.}
     \label{fig:fit-results}
\end{figure}

\begin{figure}
    \centering
    \includegraphics[width=\linewidth]{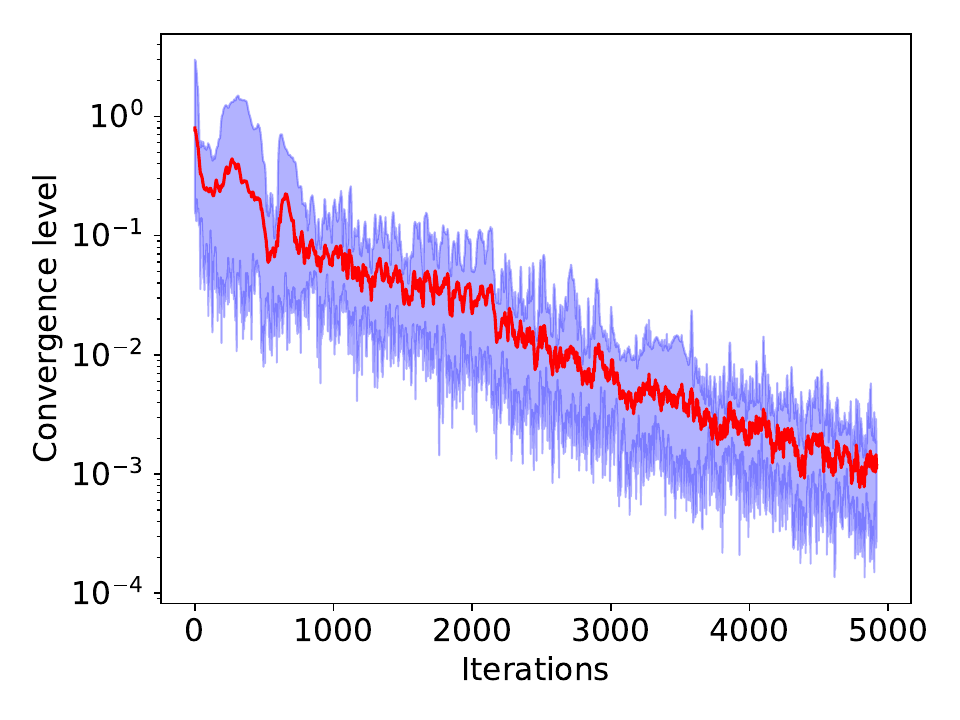}
    \caption{For each 6 parameter fit, the convergence level is defined as the maximum relative distance to the corresponding target value across all the parameters. The red line shows the average convergence level per iteration across the 10 fits. The band boundaries represent the minimum and maximum convergence level per iteration across all of the fits -- all 10 convergence levels fall within the band.}
    \label{fig:bandplot}
\end{figure}

In Fig.~\ref{fig:fit-results}, we show closure test results for fits to 10 different parameter targets using the muon sample described in Section~\ref{sec:data}. Each fit is labeled by a different color, and convergence of the 6D simultaneous fit for each parameter is shown in a separate panel. 
The targets cover a wide range of phase space. All fits converge well to their corresponding targets.
Fig.~\ref{fig:bandplot} shows a combined convergence metric across all 6 parameters, with the distribution calculated across the same 10 fits as in Fig.~\ref{fig:fit-results}. 
For each fit, the convergence level is defined as the maximum over all parameters of the relative distance of each parameter to its corresponding target value.
The band shows the maximum and the minimum relative distances to the targets among the 10 fits. All fits converge to within 1~\% of their target values after 5000 mini-batch iterations.

\begin{figure}
     \centering
     \subfloat[Charge recombination parameter $A_{B}$]{
         \centering
         \includegraphics[width=0.49\columnwidth]{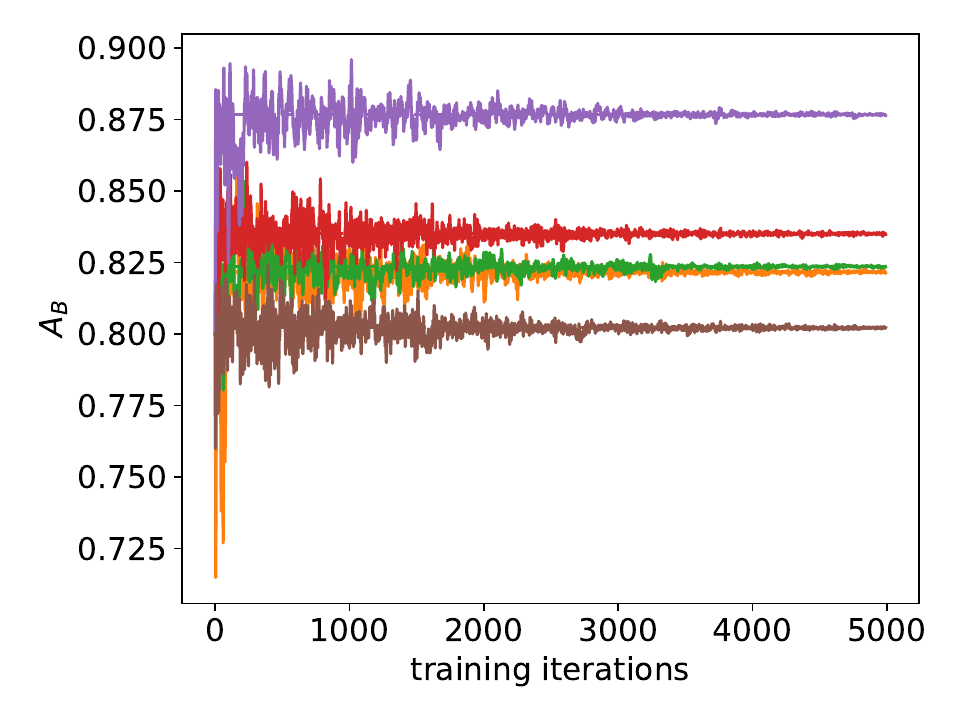}
         \label{fig:fit_result_Ab_mix}}
     \subfloat[Charge recombination parameter $k_{B}$]{
         \centering
         \includegraphics[width=0.49\columnwidth]{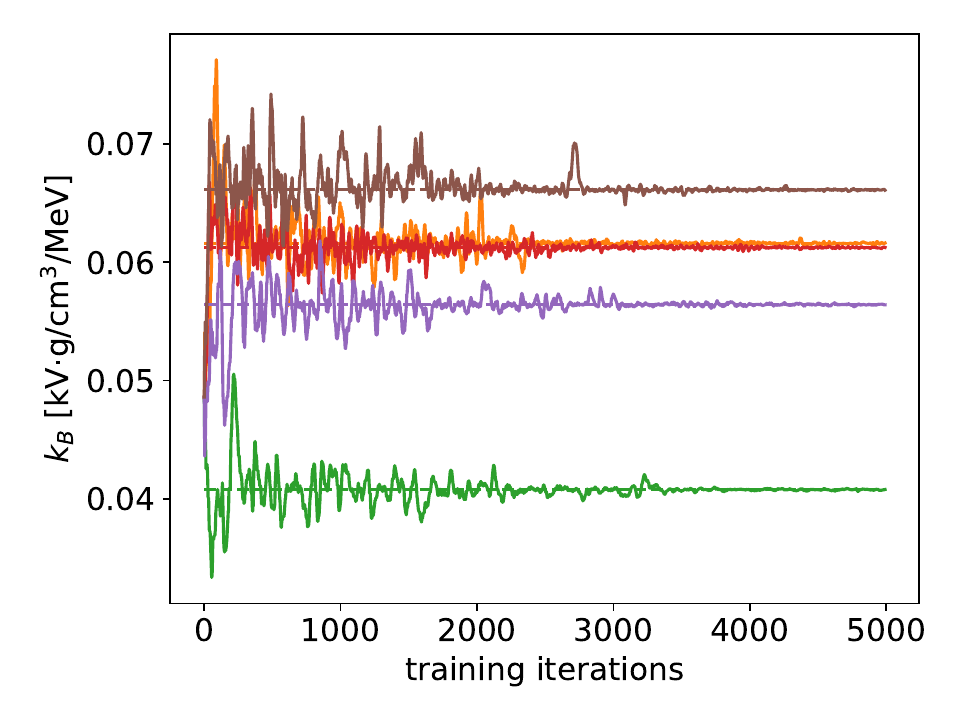}
         \label{fig:fit_result_kb_mix}}
         
     \subfloat[Drift electric field $\mathcal{E}$]{
         \centering
         \includegraphics[width=0.49\columnwidth]{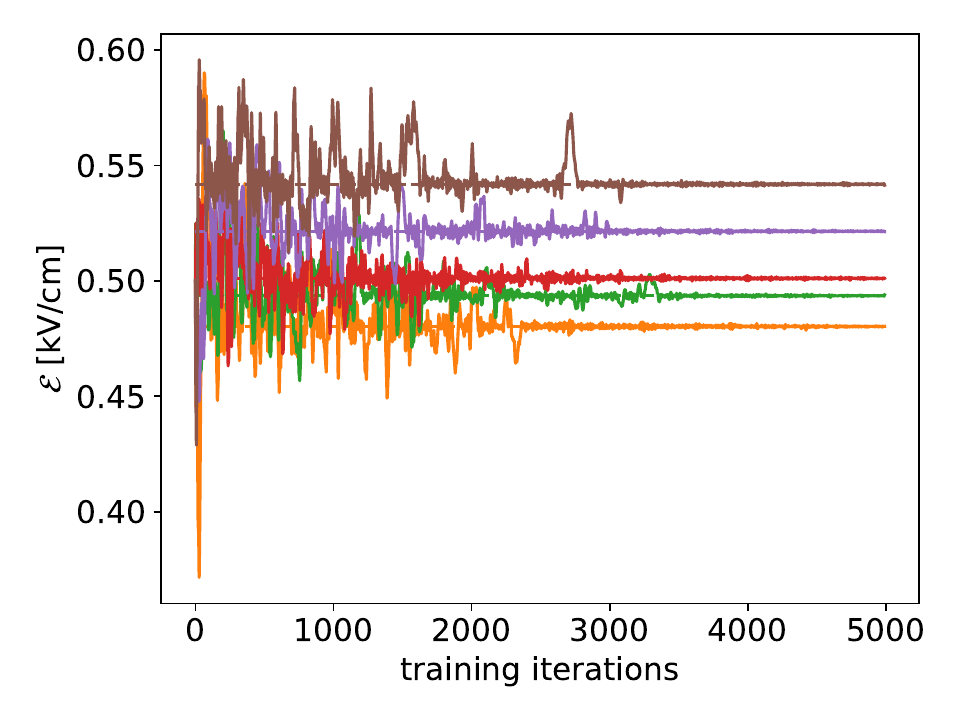}
         \label{fig:fit_result_eField_mix}}
     \subfloat[Electron lifetime $\tau$]{
         \centering
         \includegraphics[width=0.49\columnwidth]{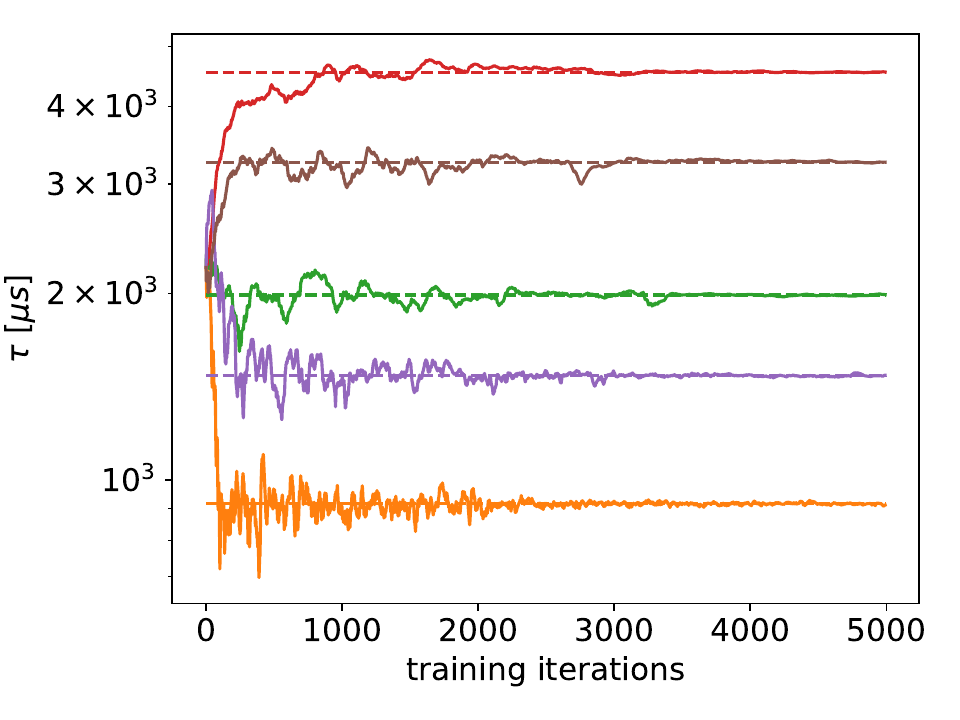}
         \label{fig:fit_result_lifetime_mix}}
         
     \subfloat[Transversal diffusion parameter $D_{T}$]{
         \centering
         \includegraphics[width=0.49\columnwidth]{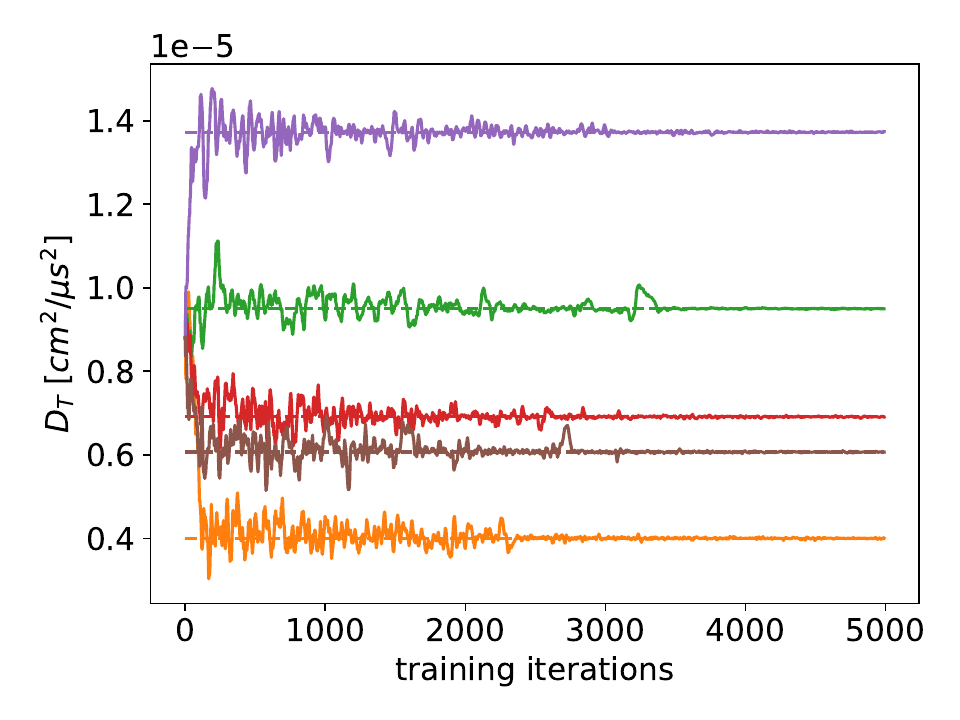}
         \label{fig:fit_result_tran_diff_mix}}
     \subfloat[Longitudinal diffusion parameter $D_{L}$]{
         \centering
         \includegraphics[width=0.49\columnwidth]{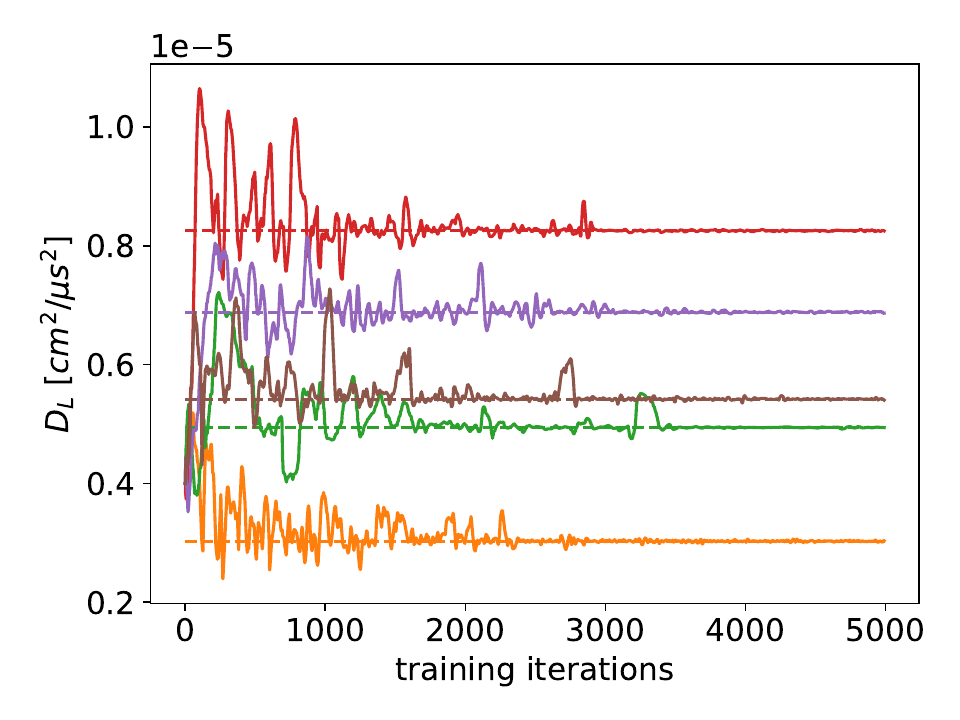}
         \label{fig:fit_result_long_diff_mix}}
     \caption{Results from a simultaneous fit of six detector model parameters: $A_{B}$, $k_{B}$, $\mathcal{E}$, $\tau$, $D_{T}$, $D_{L}$ using an alternative sample with mixed muons, pions, and protons. Each color indicates a sampled six dimensional target parameter point, and all fits start from the same initial parameter guess, mimicking a realistic fitting scenario. The dashed lines label the target values of each parameter for each respective fit. The solid lines show the evolution of fitted parameter values with respect to the iteration number in the fit.}
     \label{fig:fit-results-mixed-sample}
\end{figure}

An additional 5 fits are run using the mixed particle sample described in Sec.~\ref{sec:data}.
These fits follow the same closure test procedure as the muon sample, again taking randomly sampled parameter targets and initial values from Table~\ref{table:param_ranges}.
The mixed particle fit results are shown in Fig.~\ref{fig:fit-results-mixed-sample}, where each fit is labeled by a different color, and convergence of the 6D simultaneous fit for each parameter is shown in a separate panel.
Results are comparable to Fig.~\ref{fig:fit-results}.
This demonstrates that optimization with the differentiable simulator does not heavily depend on the particle types and energies in the input sample. This is different from conventional calibration, and suggests that our method may introduce significant flexibility in calibration fits.

The results above demonstrate that a robust calibration can be performed in a multi-dimensional phase space using gradient-based optimization. 
This novel technique enables us to simultaneously optimize a set of model parameters across the detector simulation. In addition, the number of particles required for this calibration would have a negligible impact on data-taking, and could be collected in a very short amount of time. This would allow for frequent verification of the calibration. Furthermore, this procedure is extensible to larger or different sets of parameters, making it suitable for generic use.
The optimized model parameters can be immediately applied within the same simulation used for calibration. This benefits experiments by automatically ensuring consistent application across the simulation and analysis chain.

While we have focused on the calibration aspect of the differentiable simulator, the success of the simulator in this area also provides a validation of the resulting simulator gradients. 
The ability to calculate gradients through the simulation provides opportunities for a variety of extensions. 
Highlighting one example, with a standard, non-differentiable simulator, training of neural networks operates on fixed simulation samples -- as the network trains using gradients, it cannot be directly linked to the simulation, but rather must learn what the simulation is doing from simulation outputs and inputs. In contrast, with a differentiable simulator, the simulator may be directly linked with the neural network -- gradients are able to propagate through the simulation code for learned network parameter updates, providing a direct physical constraint on learned models. Application of such a structure to learning an ``inverse detector simulation'' which maps from detector outputs to $\frac{dE}{dx}$ segments is one particular topic for future work.


\section{Conclusions and outlook}
We have presented a novel detector calibration technique for LArTPCs using differentiable detector simulation. Our simulator sets a new paradigm for a physics-model based, high dimensional, automatic calibration, allowing for direct extraction of physics information from calibration fits, trivial application of calibrated information via the setting of simulation parameters, and full accounting of all relationships between parameters in a previously inaccessible dimensionality of parameter space. This simulator is based on the configuration of a DUNE LArTPC prototype. However, we expect much of this work to be applicable beyond this particular detector configuration.

Our simulator has been demonstrated to robustly fit targets across a wide range of parameter space using multiple physics samples, and therefore provides a strong proof-of-concept demonstration of the utility of differentiable detector simulation for the calibration task. We have identified a variety of areas for future work to bring this idea from proof-of-concept to integral part of the operation of experiments using LArTPCs. This future work can be summarized in two categories:
\begin{enumerate}
\item \textbf{Physics considerations.} The presented calibration results are simulation closure tests.
In this setup, we use the exact same particle segments when generating the target data and running the calibration fit.
With real data, we will have to estimate the particle segments event-by-event.
For track-like particle samples, such as muons, pions, and protons, this can be done by fitting the readout hits with lines and breaking those lines into segments. 
We will then need to reconstruct $\frac{dE}{dx}$ using the charge readout and the reconstructed segment length. However, this reconstruction has some dependence on detector parameters, and will likely require iterative updates of $\frac{dE}{dx}$ alongside the parameter optimization fit.
Furthermore, there will likely be some associated degradation in fit quality due to a mismatch between reconstructed segments and the true particle energy deposition. This procedure needs to be studied and validated.

We have shown the impact of electronics noise on simulation output with respect to a range of parameter values in Fig.~\ref{fig:spider-plot}. This noise may degrade the performance of parameter optimization, and should be characterized for application of our calibration in experiments.

We have applied a variety of cuts to create a curated calibration sample. These cuts are on true quantities known from the simulation record, but will have to be inferred in real data via a reconstruction. 

We have presented results for the most commonly calibrated parameters. We expect our method to be easily extensible to larger parameter sets, but this should be explicitly evaluated for specific cases of interest.

A variety of updates have been made to \texttt{larnd-sim} after the snapshot used as a base for our simulator. Incorporating these changes, either via updates to our simulator or a fresh rewrite of \texttt{larnd-sim}, is necessary for broader adoption within DUNE. \texttt{larnd-sim} includes non-differentiable structures such as lookup tables, and differentiable parametrizations as in Ref.~\cite{SIREN} may be necessary. Unifying this LArTPC simulation with work on differentiable photon propagation as in Ref.~\cite{SIREN} is an additional area of development.

\item \textbf{Computational considerations.} We have identified two areas of software optimization which may decrease the computational requirements of our simulator, inspired by the computational gap between our simulator and \texttt{larnd-sim}. These include keeping a notion of sparsity in our simulator, as opposed to the current vectorized code using dense tensors, and introducing JIT compilation. JIT compilation tools are available in the supported backends for EagerPy, and exploration of these tools may prove fruitful. Pure comparison of these backends may also offer performance gains relative to our choice of PyTorch.

Though sparse libraries exist for PyTorch, JAX, and TensorFlow, moving beyond libraries designed around tensor operations may be promising. This is enabled by, for example, Enzyme~\cite{Enzyme} and is baked into the structure of non-Pythonic languages such as Julia~\cite{Julia}, which also automatically includes JIT compilation. 

In addition to optimizing computational performance, we have pointed out that there are areas of our code which remain non-differentiable, such as the discrete pixel structure. We expect that the parts that are differentiable will be sufficient for most applications. However efforts to incorporate differentiability in the pixel plane (e.g. via relaxation with kernel density estimation~\cite{neos}) may be interesting for some applications.

\end{enumerate}

The calibration tests presented here are also a test of the validity of parameter gradients through our simulator. This work therefore further sets the stage for the use of differentiable detector simulation within a broader machine learning context, allowing for e.g. explicit feedback of detector simulation on neural network training, a rich area with many applications, including learning to remove detector effects. 

In summary, our work is a first step towards a broader differentiable physics program in particle physics. This differentiable physics program has potential for significant impact in the way physics analysis is performed, and there is a broad set of interesting future tasks towards integrating differentiable toolkits within particle physics to expand analysis capability and improve the quality and output of new physics results.


\section{Acknowledgements}
The authors would like to acknowledge support from the SLAC Machine Learning Initiative. Computational resources and support were provided by the SLAC Shared
Scientific Data Facility. The authors would also like to thank Mark Convery, Dan Douglas, Francois Drielsma, and Hirohisa A. Tanaka for useful comments during paper review.

\bibliography{references}

\end{document}